\pdfoutput=1

\documentclass[a4paper]{article}
\usepackage{a4wide}
\usepackage{graphicx}
\usepackage[T1]{fontenc}
\usepackage[utf8]{inputenc}
\usepackage[UKenglish]{babel}
\usepackage{placeins}
\usepackage[shortlabels]{enumitem}
\usepackage[binary-units=true]{siunitx}
\usepackage{color}
\usepackage{authblk}
\usepackage{amsmath}
\usepackage{amssymb}
\usepackage{amsthm}
\usepackage{mathtools}
\usepackage{bbold}
\usepackage{dsfont}
\usepackage{braket}
\usepackage{tikz}
\usepackage[ruled,lined]{algorithm2e}
\usepackage{nicefrac}
\usepackage[backend=bibtex,style=phys,sorting=none,citestyle=numeric-comp,eprint=true]{biblatex}
\usepackage{hyperref}
\usepackage{cleveref}

\graphicspath{{figures/},{data/}}

\crefformat{footnote}{#2\footnotemark[#1]#3}

\DeclareMathOperator{\ord}{\mathcal{O}}

\newcommand{\eto}[1]{\ensuremath{\mathrm{e}^{#1}}}

\newcommand{\md}{\ensuremath{\mathrm{d}}}

\newcommand{\tint}{\ensuremath{\tau_\text{int}}}
\newcommand{\ordnung}[1]{\ensuremath{\ord\left(#1\right)}}
\newcommand{\erwartung}[1]{\ensuremath{\left\langle#1\right\rangle}}

\newcommand{\intaut}{\hyperref[def:tau_int]{integrated autocorrelation time}}
\newcommand{\radup}{\hyperref[def:radial_update]{radial update}}
\newcommand{\Radup}{\hyperref[def:radial_update]{Radial update}}
\newcommand{\radman}{\hyperref[def:radial_manifold]{radial Riemannian manifold}}
\newcommand{\effpot}{\hyperref[def:eff_pot]{effective potential}}
\newcommand{\approach}{\hyperref[def:approach]{approach}}

\theoremstyle{definition}
\newtheorem{definition}{Definition}[section]

\newtheorem{lemma}[definition]{Lemma}
\newtheorem{estimate}[definition]{Estimate}
\newtheorem{theorem}{Theorem}
\newtheorem{corollary}[definition]{Corollary}
\newtheorem{proposition}[definition]{Proposition}
\newtheorem{conjecture}[definition]{Conjecture}

\theoremstyle{remark}
\newtheorem*{remark}{Remark}
\newtheorem*{argument}{Argument}

\newcommand{\bonn}{
	\textit{\footnotesize Helmholtz-Institut f\"{u}r Strahlen- und
		Kernphysik,
		University of Bonn, 53115 Bonn, Germany}
}

\makeatletter%
\makeatletter%
\begin{document}
	
	\title{\vspace*{-\baselineskip}Exponential speed up in Monte Carlo sampling through Radial Updates}
	
	\author{Johann Ostmeyer\footnote{\href{mailto:ostmeyer@hiskp.uni-bonn.de}{ostmeyer@hiskp.uni-bonn.de}}}
	\affil{\bonn}
	\date{\vspace*{-2\baselineskip}}
	\maketitle
	
	\begin{abstract}
		Recently, it has been shown that the hybrid Monte Carlo (HMC) algorithm is guaranteed to converge exponentially to a given target probability distribution $p(x)\propto\eto{-V(x)}$ on non-compact spaces if augmented by an appropriate radial update. In this work we present a simple way to derive efficient radial updates meeting the necessary requirements for any potential $V$. We reduce the problem to finding a substitution for the radial direction $||x||=f(z)$ so that the effective potential $V(f(z))$ grows exponentially with $z\rightarrow\pm\infty$. Any additive update of $z$ then leads to the desired convergence. We show that choosing this update from a normal distribution with standard deviation $\sigma\approx 1/\sqrt{d}$ in $d$ dimensions yields very good results. We further generalise the previous results on radial updates to a wide class of Markov chain Monte Carlo (MCMC) algorithms beyond the HMC and we quantify the convergence behaviour of MCMC algorithms with badly chosen radial update. Finally, we apply the radial update to the sampling of heavy-tailed distributions and achieve a speed up of many orders of magnitude.
	\end{abstract}

	\unitlength = 1em
	
	\section{Introduction}

Typically, when a new Monte Carlo method is introduced, it is shown to reproduce the desired stationary probability distribution. However, the convergence towards this probability distribution, also called thermalisation process, is rarely considered in detail beyond empirical evidence. It turns out that up to the very recent past it had not been shown for algorithms as prominent as the hybrid Monte Carlo (HMC)~\cite{Duane1987} whether they converge at all.

Only recently, in Ref.~\cite{original_radial_update} Kennedy and Yu have shown analytically that the HMC is guaranteed to converge to the correct target probability distribution on compact Riemannian manifolds, see \cref{th:conv_with_cdc_sgdc}. The same, however, is not immediately clear for non-compact manifolds. Only by introducing an additional radial update, could they further prove that the combined algorithm also converges on non-compact manifolds, see \cref{th:hmc_convergence_non-compact}.

For polynomial and single-logarithmic potentials (i.e.\ no leading terms like $\log\log x$) they also provide update schemes that guarantee convergence explicitly. These updates are discrete and contain a number of free parameters. The choice of these parameters can significantly influence the performance of the update scheme and it is not clear a priori how best to choose them. For applications in practice it is therefore of high interest to derive an updating scheme that guarantees convergence, allows efficient updates, and does not require too much tuning. This work derives exactly this.

More specifically, the main results of this work are two-fold. First, we derive a general way to construct radial updates for arbitrary potentials (\cref{th:dbl_exp_subst}) and prove that this construction leads to exponential convergence. Subsequently, in \cref{th:scaling_sigma} we highlight a particular choice among the valid update schemes and show that it leads to near-optimal autocorrelation. That is, this choice provides the best coefficient in the exponential convergence achievable without case-by-case tuning.

The radial updates derived below have to be combined with other algorithms like the HMC in order to achieve optimal convergence. For the hands-on implementation of the HMC, we refer to Refs.~\cite{KENNEDY1991118,Riemann_Manifold:2011,Neal:2011mrf,NUTS-2014,apers2022hamiltonian,Ostmeyer:2024amv} where multiple improvements to the original algorithm~\cite{Duane1987} have been introduced over the years. Furthermore the convergence of the HMC has been discussed extensively prior to Ref.~\cite{original_radial_update}, for instance in Refs.~\cite{durmus2019convergencehamiltonianmontecarlo,Livingstone:2016,Livingstone:2017}.

Since Ref.~\cite{original_radial_update} initially considered the HMC and this algorithm is so prominent in physics, we will regularly remark on how our general results apply to the HMC in particular. We emphasise, however, that this work tackles a much more general question, namely how to combine any algorithm (not just the HMC) viable on compact spaces with radial updates so that the combined method converges exponentially on non-compact spaces. We will not assume any particular algorithm to be augmented.

While this manuscript was still in preparation, the combined algorithm of HMC with radial updates has already been successfully applied to a physical problem, namely the Hubbard model, for the first time~\cite{Temmen:2024pcm,Temmen_ergodicity}. It turned out that radial updates not only help with the convergence, they also allow to jump over some potential barriers. In the case at hand, this feature sufficed to remove an otherwise severe ergodicity violation.

The rest of this work is structured as follows. The radial updates are introduced and explained on an intuitive level in \cref{sec:algorithm}. Therein we also provide explicit algorithms as well as some numerical examples. In \cref{sec:formalism} the algorithm is derived, its convergence is proven, and the free parameter is tuned. We conclude in \cref{sec:conclusion}. 	
	\section{The algorithm}\label{sec:algorithm}

Most Markov chain Monte Carlo (MCMC) methods rely on local updates of limited step size, or at least quickly decaying probabilities for large steps. In principle, the local steps are a desired feature because they keep the Markov chain within a region of high probability. If an accept/reject step is used, too drastic updates result in a very low acceptance and thus inefficient sampling.

However, the limited step size is also a major weakness of such methods if either the target probability distribution has vast regions (much larger than the step size) with significant probability densities, or the initial configuration happens to be far away from the region of high probability density. In both cases, the small update steps might take prohibitively long to sample the true probability distribution reliably. \Radup s can substantially alleviate or even entirely resolve both problems.

In this section the \radup\ is introduced on an intuitive level, the \cref{alg:poly_pot,alg:any_pot} summarise the hands-on realisation, and some numerical examples are provided in \cref{sec:examples}. All the technical details including formal definitions, theorems and proofs can be found in \cref{sec:formalism,sec:proofs,sec:slow_conv}.

\subsection{Basic idea}

\begin{algorithm*}[tb]
	\caption{Radial update sampling $x\in X$ from the probability distribution $p(x)\propto\eto{-V(x)}$ generated by a polynomial potential $V$ on the Euclidean space $X=\mathds{R}^d$ based on \cref{th:updates_in_practice} (choice of log-normal update) and \cref{th:scaling_sigma} (default choice of variance $\sigma^2$). This update should be combined with some other algorithm (e.g.\ the HMC) that allows to sample the angular component correctly.}\label{alg:poly_pot}
	\SetKwInOut{Input}{input}
	\SetKwInOut{Params}{parameters}
	\SetKwInOut{Output}{output}
	\Params{dimension $d$, potential $V$ with $V(x)\approx c|x|^a$ for large $|x|$}
	\Input{initial configuration $x^\text{i}\in X$, standard deviation $\sigma$ (default $\sigma = \sqrt{\frac{2}{ad}}$)}
	\Output{final configuration $x^\text{f}\in X$}
	sample $\gamma\sim \mathcal{N}(0,\sigma^2)$ \tcp*{normal distribution}
	$x \gets x^\text{i}\cdot \eto{\gamma}$\;
	$\Delta V \gets V(x)-V(x^\text{i})$\;
	\uIf(\tcp*[f]{uniform distribution}){$\eto{-\Delta V+d\gamma}\ge \mathcal{U}_{[0,1]}$}{$x^\text{f} \gets x$\;}
	\Else{$x^\text{f} \gets x^\text{i}$\;}
\end{algorithm*}

\Radup s can be applied when the target probability distribution is defined on some non-compact space and each point within this space has a well-defined norm or radius. Then this radius can be updated independently of the angular components and the step size can be scaled with the current value of the radius. This means that remote regions with large radii can be left quickly, but also that far-away regions with high probability can be approached quickly.

The idea behind \radup s is not to replace other sampling techniques but to complement them. Start with your favourite sampling algorithm (e.g.\ the HMC), perform a fixed number $n_0$ of updates with this algorithm and then some fixed number $n_r$ of \radup s. The combined algorithm is designed to mitigate the weaknesses of both components. The two integers $n_0$ and $n_r$ can be chosen freely and should be tuned depending on the case at hand. That said, $n_0=n_r=1$ is typically a good starting point.

In such a combined updating scheme, the initial algorithm is responsible for sampling the local environment and in particular the angular component of the configurations. The \radup\ on the other hand guarantees reliable sampling of the radial component (as the name suggests), even if this requires jumps across several orders of magnitude. Without \radup s most classical MCMC algorithms will enter an inefficient diffusive regime in low-probability regions if the probability density does not decay quickly enough.
After a local update, the probability density hardly changes because it appears similarly small in the entire region. Thus, the algorithm becomes insensitive to the probability distribution. In extreme cases, this could lead to the simulation wandering off towards infinity.
This effect is negated by the \radup\ because its large steps in remote areas remain sensitive to changes of the probability density.

This sensitivity can only be guaranteed if the asymptotic scaling of the probability density is known. For instance, the most common case in physics is a probability distribution $p(x)\propto \eto{-V(x)}$ defined by an asymptotically polynomial potential, e.g.\ $V(x)=c|x|^2+\ordnung{x}$. The \radup\ required in this case has been summarised in \cref{alg:poly_pot}.

More generally, the \radup\ can be defined using a substitution. The reason is that even a local update is sensitive to changes in the probability density if said density decays double-exponentially, i.e.\ the potential grows exponentially. Thus, if the radius $r=||x||$ is substituted by $r=f(z)$ so that the \effpot\ in the auxiliary variable $z$ grows exponentially $V(f(z))\approx c\eto{a|z|}$, then the \radup\ can simply choose local additive steps for $z$. Intuitively, additive updates in the exponent lead to multiplicative updates of the potential and repeated multiplicative updates imply exponential convergence. For the polynomial potential in the example above a sensible substitution is $r=\eto{z}$. \Cref{alg:any_pot} details the generalised \radup\ relying on an appropriate substitution.

\begin{algorithm*}[tb]
	\caption{Radial update sampling $x\in X$ from the probability distribution $p(x)\propto\eto{-V(x)}$ generated by an arbitrary potential $V$ on the \radman\ $X$ based on \cref{th:dbl_exp_subst}. This update should be combined with some other algorithm (e.g.\ the HMC) that allows to sample the angular component $\theta$ correctly.}\label{alg:any_pot}
	\SetKwInOut{Input}{input}
	\SetKwInOut{Params}{parameters}
	\SetKwInOut{Output}{output}
	\Params{dimension $d$, potential $V$}
	\Input{initial configuration $x^\text{i}\in X$, standard deviation $\sigma$, substitution $f$ with Jacobian $J$ so that $V_\text{eff}(z,\theta)\coloneqq V(f(z),\theta) - \ln\left|\det J(z,\theta)\right|= c(\theta)\eto{a(\theta) |z|+o(|z|)}$}
	\Output{final configuration $x^\text{f}\in X$}
	$z^\text{i} \gets f^{-1}(||x^\text{i}||)$ \tcp*{rewrite $x^\text{i}=(f(z^\text{i}),\theta)$}
	sample $\gamma\sim \mathcal{N}(0,\sigma^2)$ \tcp*{normal distribution}
	$z \gets z^\text{i} + \gamma$\;
	$x \gets (f(z),\theta)$\;
	$\Delta V \gets V_\text{eff}(x)-V_\text{eff}(x^\text{i})$\;
	\uIf(\tcp*[f]{uniform distribution}){$\eto{-\Delta V}\ge \mathcal{U}_{[0,1]}$}{$x^\text{f} \gets x$\;}
	\Else{$x^\text{f} \gets x^\text{i}$\;}
\end{algorithm*}

In addition to the choice of substitution, the \radup\ allows for any local additive update for $z$. Originally, in Ref.~\cite{original_radial_update} discrete updates were proposed. This is a valid but not very efficient choice. In fact, one of the main results of this work is that for maximal efficiency (or minimal \intaut) the update should be proposed from a normal distribution and for polynomial potentials the standard deviation $\sigma$ should be scaled as the inverse square root of the dimension $\sigma\propto1/\sqrt{d}$. The proportionality coefficient in this scaling is case-dependent, but it can be estimated up to deviations of order one. The best estimator that typically also performs very well in practice, has been added as the default value in \cref{alg:poly_pot}.

When tuning $\sigma$ empirically, a good target acceptance rate is around $50\%$. Note that it is advisable to choose $\sigma$ slightly too large rather than too small. Larger standard deviations merely reduce the acceptance rate linearly while smaller $\sigma$ lead to a diffusive regime in which autocorrelations grow quadratically.

\subsection{Applications and numerical examples}\label{sec:examples}

The \radup\ has been designed specifically to help with the sampling from heavy-tailed distributions generated by slowly growing potentials. Here we provide some numerical examples demonstrating that a correctly chosen \radup\ does indeed allow to sample from otherwise intractable distributions.

\begin{figure}[t]
	\centering

 	\caption{Histograms of the Markov chains ($10^5$ steps) generated using \cref{alg:poly_pot,alg:any_pot}, respectively, based on \cref{th:dbl_exp_subst,th:updates_in_practice,th:scaling_sigma}. The potential $V(x)$ depends only on the radius $|x|\equiv r=f(z)$ and the angular component was not sampled for simplicity. In both cases $z\in \mathds{R}$ was updated by $z\rightarrow z+\gamma$ with $\gamma\sim\mathcal{N}(0,\frac 2d)$. Left: $d=100$ dimensions, $V(x)=|x|$ (i.e.\ $p(r)\propto r^{d-1}\eto{-r}$) with the substitution $r=\eto{z}$ and \effpot\ $V_\text{eff}(z)=\eto{z}-dz$; Right: $d=1$ dimension, $V(x)=\ln\left(1+|x|^{\num{1.01}}\right)$ (i.e.\ $p(r)\propto\frac{1}{1+r^{\num{1.01}}}$) with the substitution $r=\eto{\sinh z}$ and $V_\text{eff}(z)=\ln\left(1+\eto{\num{1.01}\sinh z}\right)-\sinh z - \ln\cosh z$.}\label{fig:histograms}
\end{figure}

Figure~\ref{fig:histograms} shows the histograms of two probability distributions sampled with \radup s. The substitutions for the respective \radup\ were chosen according to \cref{th:updates_in_practice,th:scaling_sigma} as summarised in \cref{alg:poly_pot} for the polynomial potential and in \cref{alg:any_pot} for general ones. On the left hand side the target potential is $V(r)=r$ in $d=100$ dimensions and on the right $V(r)=\ln\left(1+r^{\num{1.01}}\right)$ with $d=1$ (keep in mind that $V(r)=\ln\left(1+r\right)$ would not be normalisable). Within statistical fluctuations both histograms reproduce the target distribution.
Note the logarithmic scale in the right panel of figure~\ref{fig:histograms} with significant probability densities for radii as large as $r=10^{200}$, all sampled reliably.

\begin{figure}[th]
	\centering
	\resizebox{0.58\textwidth}{!}{{\large%
\begingroup
  \inputencoding{latin1}%
  \makeatletter
  \providecommand\color[2][]{%
    \GenericError{(gnuplot) \space\space\space\@spaces}{%
      Package color not loaded in conjunction with
      terminal option `colourtext'%
    }{See the gnuplot documentation for explanation.%
    }{Either use 'blacktext' in gnuplot or load the package
      color.sty in LaTeX.}%
    \renewcommand\color[2][]{}%
  }%
  \providecommand\includegraphics[2][]{%
    \GenericError{(gnuplot) \space\space\space\@spaces}{%
      Package graphicx or graphics not loaded%
    }{See the gnuplot documentation for explanation.%
    }{The gnuplot epslatex terminal needs graphicx.sty or graphics.sty.}%
    \renewcommand\includegraphics[2][]{}%
  }%
  \providecommand\rotatebox[2]{#2}%
  \@ifundefined{ifGPcolor}{%
    \newif\ifGPcolor
    \GPcolortrue
  }{}%
  \@ifundefined{ifGPblacktext}{%
    \newif\ifGPblacktext
    \GPblacktexttrue
  }{}%
  \let\gplgaddtomacro\g@addto@macro
  \gdef\gplbacktext{}%
  \gdef\gplfronttext{}%
  \makeatother
  \ifGPblacktext
    \def\colorrgb#1{}%
    \def\colorgray#1{}%
  \else
    \ifGPcolor
      \def\colorrgb#1{\color[rgb]{#1}}%
      \def\colorgray#1{\color[gray]{#1}}%
      \expandafter\def\csname LTw\endcsname{\color{white}}%
      \expandafter\def\csname LTb\endcsname{\color{black}}%
      \expandafter\def\csname LTa\endcsname{\color{black}}%
      \expandafter\def\csname LT0\endcsname{\color[rgb]{1,0,0}}%
      \expandafter\def\csname LT1\endcsname{\color[rgb]{0,1,0}}%
      \expandafter\def\csname LT2\endcsname{\color[rgb]{0,0,1}}%
      \expandafter\def\csname LT3\endcsname{\color[rgb]{1,0,1}}%
      \expandafter\def\csname LT4\endcsname{\color[rgb]{0,1,1}}%
      \expandafter\def\csname LT5\endcsname{\color[rgb]{1,1,0}}%
      \expandafter\def\csname LT6\endcsname{\color[rgb]{0,0,0}}%
      \expandafter\def\csname LT7\endcsname{\color[rgb]{1,0.3,0}}%
      \expandafter\def\csname LT8\endcsname{\color[rgb]{0.5,0.5,0.5}}%
    \else
      \def\colorrgb#1{\color{black}}%
      \def\colorgray#1{\color[gray]{#1}}%
      \expandafter\def\csname LTw\endcsname{\color{white}}%
      \expandafter\def\csname LTb\endcsname{\color{black}}%
      \expandafter\def\csname LTa\endcsname{\color{black}}%
      \expandafter\def\csname LT0\endcsname{\color{black}}%
      \expandafter\def\csname LT1\endcsname{\color{black}}%
      \expandafter\def\csname LT2\endcsname{\color{black}}%
      \expandafter\def\csname LT3\endcsname{\color{black}}%
      \expandafter\def\csname LT4\endcsname{\color{black}}%
      \expandafter\def\csname LT5\endcsname{\color{black}}%
      \expandafter\def\csname LT6\endcsname{\color{black}}%
      \expandafter\def\csname LT7\endcsname{\color{black}}%
      \expandafter\def\csname LT8\endcsname{\color{black}}%
    \fi
  \fi
    \setlength{\unitlength}{0.0500bp}%
    \ifx\gptboxheight\undefined%
      \newlength{\gptboxheight}%
      \newlength{\gptboxwidth}%
      \newsavebox{\gptboxtext}%
    \fi%
    \setlength{\fboxrule}{0.5pt}%
    \setlength{\fboxsep}{1pt}%
    \definecolor{tbcol}{rgb}{1,1,1}%

}}
	\caption{Time series (left) and histogram (right) of the Markov chains ($\num{3e5}$ steps) generated using additive updates $r\rightarrow r+\gamma$ with $\gamma\sim\mathcal{N}(0,2)$ without appropriate \radup. The radial potential $V(r)=\ln\left(1+r^{\num{1.1}}\right)$ (i.e.\ $p(r)\propto\frac{1}{1+r^{\num{1.1}}}$) of the radius $r\ge 0$ was considered in $d=1$ dimension. The different time series correspond to different random number seeds.}\label{fig:ts_hist_lin}
\end{figure}

As a comparison, the less heavy-tailed distribution generated by $V(r)=\ln\left(1+r^{\num{1.1}}\right)$ in $d=1$ dimension was also attempted to sample using purely additive updates without substitution. The results can be seen in figure~\ref{fig:ts_hist_lin}. Already from the time series on the left the very long autocorrelation times are evident. The histogram on the right shows that the distribution is not even remotely correctly sampled. In order to visit the region around $r=10^{200}$ as in the previous example with this update scheme, one would have to sample over times $t\gg 10^{400}$ because it becomes an almost diffusive process in the flat regions of the probability distribution. This means that for any practical purpose such a sampler without appropriate \radup\ is not merely inefficient, it will simply provide wrong results.  	
	\section{Formal derivation}\label{sec:formalism}

This section formalises the concepts introduced in \cref{sec:algorithm}. Any reader who is interested in the practical application of the algorithm and not in the mathematical details, can safely skip this section.

All the proofs that are left out in this section, have been moved to \cref{sec:proofs}, or they are readily available in the literature (the corresponding reference is then explicitly cited within the theorem). Only very short proofs and those that are both new and insightful in their own right are given in the main manuscript.

This section is divided into three parts. In \cref{sec:radup_exp} the main results of this work are derived, leading up to \cref{th:dbl_exp_subst,th:updates_in_practice} which describe how to choose a \radup\ that guarantees exponential convergence to the desired probability distribution. These results are generalised to less efficient but still convergent update schemes in \cref{sec:general_conv}. Finally, in \cref{sec:param_tuning} we derive the optimal choice of the standard deviation in a \radup\ applied to (asymptotically) polynomial potentials.

\subsection{Radial update and exponential convergence}\label{sec:radup_exp}
Throughout this manuscript we use the standard Landau symbols as in \cref{tab:landau_symbols}. We also use the convention $|\cdot|\equiv||\cdot||_2$, that is the norm with single bars always denotes the standard Euclidean norm.

\begin{table}
	\centering
	\begin{tabular}{c|l|c}
		Notation & \multicolumn{1}{c|}{Meaning} & Definition \\\hline
		$f\in o(g)$ & $f$ grows strictly slower than $g$ & $\lim\limits_{x\rightarrow x_0}\left|\frac{f(x)}{g(x)}\right|=0$\\
		$f\in \ordnung{g}$ & $f$ grows at most as quickly as $g$ & $\limsup\limits_{x\rightarrow x_0}\left|\frac{f(x)}{g(x)}\right|<\infty$\\
		$f\in \Theta(g)$ & $f$ grows exactly as quickly as $g$ & $0<\liminf\limits_{x\rightarrow x_0}\left|\frac{f(x)}{g(x)}\right|\le \limsup\limits_{x\rightarrow x_0}\left|\frac{f(x)}{g(x)}\right|<\infty$\\
		$f\in \Omega(g)$ & $f$ grows at least as quickly as $g$ & $\liminf\limits_{x\rightarrow x_0}\left|\frac{f(x)}{g(x)}\right|>0$\\
		$f\in \omega(g)$ & $f$ grows strictly faster than $g$ & $\lim\limits_{x\rightarrow x_0}\left|\frac{f(x)}{g(x)}\right|=\infty$\\
	\end{tabular}
	\caption{Standard Landau symbols (often referred to as the ``big-O-notation'') as used throughout this work. Here, $x_0$ denotes the limit of interest. If not stated otherwise, it is assumed that in spherical coordinates $x_0=(\infty,\theta_0)$ and the respective definition has to hold for all angles $\theta_0$.}\label{tab:landau_symbols}
\end{table}

\begin{definition}[Markov transition kernel~\cite{hairer2008look}]\label{def:markov_kernel}
	The Markov transition kernel $\mathcal{P}:X\times X\rightarrow[0,\infty)$ on the measurable space $X$ denotes the transition probability density $\mathcal{P}(x,y)$ from $x$ to $y$ for a given update scheme. It is normalised
	\begin{align}
		\int_X \md y \mathcal{P}(x,y) &= 1
	\end{align}
	for all $x\in X$. For a given measurable function $V : X\rightarrow[0,\infty)$, it acts as
	\begin{align}
		(\mathcal{P}V)(x) &= \int_X \md y \mathcal{P}(x,y) V(y)\,.
	\end{align}
\end{definition}
\newcommand{\mtk}{\hyperref[def:markov_kernel]{Markov transition kernel}}

\begin{remark}
	$(\mathcal{P}V)(x)$ is simply the expectation value of $V$ after a single update step starting from $x$.
\end{remark}

\begin{definition}[Metropolis-Hastings accept/reject step~\cite{Metropolis:1953,Hastings_1970}]\label{def:metropolis}
	Given a \mtk\ $\mathcal{P}$ on the measurable space $X$ and a target probability distribution $p$, let an update from $x\in X$ to $y\in X$ be proposed according to the probability density $\mathcal{P}(x,y)$. In the Metropolis-Hastings accept/reject step the old state $x$ is replaced by $y$ with the probability
	\begin{align}
		p_\text{acc} &= \min\left(1,\frac{p(y)}{p(x)}\frac{\mathcal{P}(y,x)}{\mathcal{P}(x,y)}\right)
	\end{align}
	and kept unchanged otherwise.
\end{definition}
\newcommand{\mhacc}{\hyperref[def:metropolis]{Metropolis-Hastings accept/reject step}}

\begin{remark}
	The \mhacc\ guarantees detailed balance and is practically universally used in the construction of MCMC algorithms for this reason.
\end{remark}

\begin{definition}[Compact Doeblin's condition~\cite{hairer2008look,original_radial_update,Doeblin:1938,Doeblin:1940,Doob:1948}]\label{def:compact_doeblin}
	An update scheme defined by the \mtk\ $\mathcal{P}$ satisfies the Compact Doeblin's condition (CDC) on the measurable space $X$ if for all $R>0$ with $\mathcal{C}=\left\{x\in X: V(x) \le R\right\}$ compact there exist a $\xi>0$ and a probability density $\nu$ so that for all $x,y\in \mathcal{C}$
	\begin{align}
		\mathcal{P}(x,y) \ge \xi \nu(y)\,.\label{eq:cdc}
	\end{align}
\end{definition}
\newcommand{\cdc}{\hyperref[def:compact_doeblin]{CDC}}

\begin{remark}
	The \cdc\ means that any point $y$ can be reached from any starting point $x$ with finite probability density. This can only work on compact spaces because on unbounded spaces the transition probability must go to zero in order to stay normalisable.\\
	The \cdc\ as formulated here differs from Doeblin's original condition~\cite{Doeblin:1938,Doeblin:1940} in two points. Initially $y$ was not restricted to $\mathcal{C}$, but it was shown in~\cite{original_radial_update} that $y\in\mathcal{C}$ suffices, so we use this simpler formulation. Furthermore, typically it is only required that there exists an $R$ large enough. Clearly, this formulation implies the classical one and it is easier to work with.
\end{remark}

\begin{proposition}[Gaussian update]\label{th:gauss_update}
	An \mtk\ $\mathcal{P}$ defined by $x\mapsto x + \eta$ with $\eta\sim \mathcal{N}(0,\Sigma)$ sampled from a multivariate normal distribution satisfies the \cdc.
\end{proposition}

\begin{proof}
	The normal distribution has a non-zero probability density everywhere. Since the \cdc\ only considers updates within a compact region, the probability density is guaranteed to reach a non-zero minimum within this region. This minimum can be used to define a value $\xi>0$ for equation~\eqref{eq:cdc}.
\end{proof}

\begin{definition}[Strong geometric drift condition~\cite{Harris:1955,hairer2008look}]\label{def:sgdc}
	An update scheme defined by the \mtk\ $\mathcal{P}$ satisfies the strong geometric drift condition (SGDC) for the function $V:X\rightarrow[0,\infty)$ on the measurable space $X$ if some constants $K\in[0,\infty)$ and $\alpha \in [0,1)$ exist so that for all $x\in X$
	\begin{align}
		(\mathcal{P}V)(x) &\le \alpha V(x) + K\,.
	\end{align}
\end{definition}
\newcommand{\sgdc}{\hyperref[def:sgdc]{SGDC}}
\newcommand{\wgdc}{\hyperref[def:sgdc]{WGDC}}
\newcommand{\sadc}{\hyperref[def:sadc]{SADC}}
\newcommand{\wadc}{\hyperref[def:sadc]{WADC}}
\newcommand{\ladc}{\hyperref[def:ladc]{LADC}}

\begin{remark}
	Analogously we can define the \textit{weak} geometric drift condition (WGDC) if we allow $\alpha\in[0,1]$. On non-compact Riemannian manifolds additive updates like the HMC merely satisfies the \wgdc\ in general.
\end{remark}

\begin{theorem}[Convergence of MCMC algorithms~\cite{Harris:1955,Nummelin_1984,Chan:1989,Douc:2004,hairer2008look}]\label{th:conv_with_cdc_sgdc}
	An MCMC algorithm on the measurable space $X$ with a stationary probability distribution $p(x)\propto \eto{-V(x)}$ is guaranteed to converge exponentially to said distribution if it satisfies the \cdc\ and the \sgdc\ for the potential $V$ on $X$. If $X$ is compact, the \sgdc\ is always satisfied, so the \cdc\ suffices.
\end{theorem}

\begin{definition}[Radial Riemannian manifold]\label{def:radial_manifold}
	On a finite-dimensional smooth connected and complete Riemannian manifold $X$ the radius of $x\in X$ is defined via the natural metric $g$ and a given (arbitrary) central point $x_0\in X$ as $r\equiv ||x||\coloneqq g(x_0,x)$. We call $X$ a radial Riemannian manifold if there exists a point $x_0\in X$ so that every $x\in X$ can be uniquely written as a tuple $x=(r,\theta)$ where the angular component $\theta\in\Omega$ is an element of the $(n-1)$-dimensional compact Riemannian sub-manifold $\Omega$.
\end{definition}

\begin{definition}[Radial update~\cite{original_radial_update}]\label{def:radial_update}
	On a \radman\ $X$ a radial update $X\rightarrow X$, $x=(r,\theta)\mapsto (r',\theta)$ is generated by a \mtk\ $\mathcal{P}$ with transition probability density $\mathcal{P}(r,r')$ and \mhacc\ based on some target probability density $p:X\rightarrow[0,\infty)$.
\end{definition}

\begin{remark}
	On any Euclidean space (i.e. $\mathds{R}^n$), or `nice' subsets thereof, the radius $r$ can simply be identified with the Euclidean norm  $r\equiv||x||_2$. This means that the decomposition $x=(r,\theta)$ is always possible on Euclidean spaces and thus a radial update can always be introduced.
\end{remark}

\begin{theorem}[Convergence with \radup~\cite{original_radial_update}]\label{th:hmc_convergence_non-compact}
	On a \radman\ $X$ any algorithm with the stationary probability distribution $p(x)\propto\eto{-V(x)}$ that satisfies the \cdc\ and the \wgdc\ for the potential $V$ on $X$ is guaranteed to converge exponentially to $p(x)$ if it is combined with a \radup\ and the generating \mtk\ $\mathcal{P}$ satisfies the \sgdc\ for $V$ on $X\setminus X_0$ with some compact $X_0$.
\end{theorem}

\begin{lemma}\label{th:hmc_conv_compact}
	The HMC algorithm on \radman s satisfies the \cdc\ if the potential $V$ is continuous~\cite{original_radial_update}.
\end{lemma}

\begin{corollary}[HMC convergence~\cite{original_radial_update}]\label{th:really_hmc_convergence_non-compact}
	The HMC algorithm converges on compact Riemannian manifolds for smooth potentials. Together with a \radup\ as in \cref{th:hmc_convergence_non-compact} the HMC also converges exponentially on non-compact \radman s.
\end{corollary}

\begin{remark}
	In practice this means that the HMC takes care of the angular part (because it is compact) and of some compact region $X_0$. The \radup\ on the other hand guarantees that far away and asymptotically very unlikely regions $X\setminus X_0$ are left sufficiently quickly. The pure HMC might otherwise `wander off' to infinity or at least get stuck in a suppressed region overly long. Especially in high dimensions spherically symmetric updates like the HMC have a very small expected contribution in the radial direction.
\end{remark}

\begin{definition}[Trivial kernel]\label{def:triv_ker}
	A \mtk\ $\mathcal{P}$ with $\mathcal{P}(x,y) = \delta(x-y)$ up to measure zero is called trivial. Conversely, for a non-trivial \mtk\ there exist $x\in X$, $\epsilon>0$, and $\delta>0$ so that
	\begin{align}
		\int_{B_\delta(x)} \md y \mathcal{P}(x,y) &< 1-\epsilon\,,
	\end{align}
	i.e.\ the probability of an update to leave the open $\delta$-ball $B_\delta(x)$ around $x$ is at least $\epsilon$.
\end{definition}
\newcommand{\nontriv}{\hyperref[def:triv_ker]{non-trivial}}

\begin{remark}
	A trivial kernel simply does not perform any update. It has identity transition matrix and maps every point onto itself.
\end{remark}

\begin{definition}[Symmetric additive kernel]\label{def:sym_add_ker}
	A \mtk\ $\mathcal{P}$ is called additive if the transition probability from $z\in\mathds{R}$ to $z'\in\mathds{R}$ depends only on their difference $\gamma\coloneqq z'-z$, i.e.\ $\mathcal{P}(z,z+\gamma)\equiv\rho(\gamma)$ with the probability density $\rho$ independent of $z$. Moreover, $\mathcal{P}$ is symmetric if $\mathcal{P}(z,z') = \mathcal{P}(z',z)$. Thus, a symmetric additive kernel fulfils $\mathcal{P}(z,z+\gamma) = \mathcal{P}(z,z-\gamma)$, or equivalently $\rho(\gamma)=\rho(-\gamma)$ for all $\gamma\in\mathds{R}$.
\end{definition}
\newcommand{\symker}{\hyperref[def:sym_add_ker]{symmetric additive kernel}}
\newcommand{\symaker}{\hyperref[def:sym_add_ker]{symmetric kernel}}
\newcommand{\addker}{\hyperref[def:sym_add_ker]{additive kernel}}

\begin{lemma}[Exponential potentials]\label{th:exp_pot}
	For every potential $V:\mathds{R}\rightarrow \mathds{R}$ with $V(z) = c\eto{a |z|+o(|z|)}$, $c,a>0$ every \nontriv\ \symker\ $\mathcal{P}$ together with a \mhacc\ satisfies the \sgdc\ for $V$.
\end{lemma}

\begin{remark}
	A variety of integrals is most efficiently solved numerically using double exponential substitution~\cite{dbl_exp_trafo}. This property appears to be closely related to stochastic integration that requires the \sgdc\ for exponential convergence.
	
	As will become clear in the proof, the symmetry of the \mtk\ is sufficient but not necessary. There can be asymmetric kernels satisfying the \sgdc\ as well. However, we do not expect such updates to be very useful in practice and therefore restrain ourselves to this much easier and more relevant case.
\end{remark}

\begin{proof}
	First, we assign an acceptance probability $A(z, z')$ to the Metropolis-Hastings step. Since the \mtk\ is symmetric, $A(z,z')$ simplifies to a pure Metropolis acceptance, that is the ratio of the Boltzmann weights
	\begin{align}
		A(z,z') &= \min\left[1,\exp\left(-(V(z')-V(z))\right)\right]\,.
	\end{align}
	Asymptotically for large $z$ (w.l.o.g.\ we assume $z>0$ because it only enters the potential as $|z|$) this further simplifies to
	\begin{align}
		A(z,z+\gamma) &= \begin{cases}
			1 & \text{if } \gamma \le 0\,,\\
			\exp\left(-V(z)\left(\eto{a\gamma}-1\right)\right) & \text{else,}
		\end{cases}
	\end{align}
	where we have used that sub-leading contributions summarised in the little-$o$ notation can be neglected asymptotically. The asymptotic considerations suffice because everything else can be absorbed into a compact interval $[-R,R]$ for some $R\ge0$ large enough and the \sgdc\ always holds on compact spaces.
	
	We can now calculate $(\mathcal{P}_AV)(z)$, i.e.\ an update step generated by $\mathcal{P}$ with an accept/reject step, explicitly (again $|z|>R$ asymptotically large and w.l.o.g.\ $z>0$, i.e.\ $z>R$)
	\begin{align}
		(\mathcal{P}_AV)(z) &= \int_{-\infty}^{\infty}\md z'\mathcal{P}(z,z')\left[A(z,z')V(z') + \left(1-A(z,z')\right)V(z)\right]\label{eq:def_pav_1d}\\
		&= \int_{-\infty}^{\infty}\md \gamma\mathcal{P}(z,z+\gamma)\left[A(z,z+\gamma)V(z+\gamma) + \left(1-A(z,z+\gamma)\right)V(z)\right]\label{eq:p_of_v}\\
		&= \int_{-\infty}^{\infty}\md \gamma\rho(\gamma)\left[A(z,z+\gamma)c\eto{a(z+\gamma)} + \left(1-A(z,z+\gamma)\right)c\eto{az}\right]\\	
		&= \int_{-\infty}^{\infty}\md \gamma\rho(\gamma)\left[1 + A(z,z+\gamma)\left(\eto{a\gamma}-1\right)\right]c\eto{az}\\	
		\begin{split}
			&= \int_{-\infty}^{0}\md \gamma\rho(\gamma)\left[1 + \eto{a\gamma}-1\right]V(z)\\
			&\quad+\int_{0}^{\infty}\md \gamma\rho(\gamma)\left[1 + \exp\left(-V(z)\left(\eto{a\gamma}-1\right)\right)\left(\eto{a\gamma}-1\right)\right]V(z)
		\end{split}\\	
		&= \int_{0}^{\infty}\md \gamma\rho(\gamma)\left[1 + \eto{-a\gamma} + \exp\left(-V(z)\left(\eto{a\gamma}-1\right)\right)\left(\eto{a\gamma}-1\right)\right]V(z)\\	
		&= \left(\frac12+\int_{0}^{\infty}\md \gamma \rho(\gamma)\eto{-a\gamma}\right)V(z) + \int_{0}^{\infty}\md \gamma\rho(\gamma) \eto{-V(z)\left(\eto{a\gamma}-1\right)}V(z)\left(\eto{a\gamma}-1\right)\\
		&= \alpha V(z) + K\,,
	\end{align}
	where we defined
	\begin{align}
		\alpha &\coloneqq \frac12+\int_{0}^{\infty}\md \gamma \rho(\gamma)\eto{-a\gamma}<1\,,\\
		K &\coloneqq \int_{0}^{\infty}\md \gamma\rho(\gamma) \eto{-V(z)\left(\eto{a\gamma}-1\right)}V(z)\left(\eto{a\gamma}-1\right)\le \frac{1}{2e}\,.\label{eq:def_K}
	\end{align}
	$\rho$ is a symmetric normalised probability density, i.e.\ $\int_{0}^{\infty}\md \gamma \rho(\gamma)=\frac12$, and $0<\eto{-a\gamma}<1$ for all $\gamma>0$. Together with $\mathcal{P}$ and therefore $\rho$ being \nontriv, this implies $0<\alpha<1$. Furthermore the integrand in $K$ has the form $0\le\eto{-f(\gamma)}f(\gamma)\le1/e$ which is bounded since $f(\gamma)\ge0$, thus $K$ is bounded as well.
\end{proof}

\begin{definition}[Effective potential]\label{def:eff_pot}
	Write a \radman\ as $X=[0,\infty)\times\Omega$ and let $f:\mathds{R}\rightarrow[0,\infty)$ be a diffeomorphism so that $F:\mathds{R}\times\Omega\rightarrow X$, $(z,\theta)\mapsto x=F(z,\theta)=(f(z),\theta)$ is a substitution with the Jacobian $J(z,\theta)= DF(z,\theta)$. Further, let $V$ be a potential. Then the effective potential for $V$ and $f$ is defined as
	\begin{align}
		V_\text{eff}(z,\theta) &\coloneqq V(F(z,\theta)) - \ln\left|\det J(z,\theta)\right|\,.\label{eq:def_eff_pot}
	\end{align}
	In particular, on Euclidean spaces in $d$ dimensions
	\begin{align}
		V_\text{eff}(z,\theta) &=V(f(z),\theta) - (d-1)\ln f(z) - \ln \left| f'(z)\right|\,.\label{eq:def_eucl_eff_pot}
	\end{align}
\end{definition}

Examples of \effpot s, can be found in the caption of figure~\ref{fig:histograms} for a polynomial and a logarithmic potential.

\begin{theorem}[Convergence with (double) exponential substitution]\label{th:dbl_exp_subst}
	On a \radman\ $X=[0,\infty)\times\Omega$ take any algorithm with the stationary probability distribution $p(x)\propto\eto{-V(x)}$ that satisfies the \cdc\ and the \wgdc\ for the potential $V$ on $X$. Add a \radup\ generated by an arbitrary \nontriv\ \symker\ $\mathcal{P}$ acting on $z\in\mathds{R}$ via a substitution $f$ so that the \effpot\ for $V$ and $f$ satisfies
	\begin{align}
		 V_\text{eff}(z,\theta) &= c(\theta)\eto{a(\theta) |z|+o(|z|)}\label{eq:subst_condition}
	\end{align}
	with $c(\theta),a(\theta)>0$ for every $\theta\in\Omega$. Then the combined update converges exponentially to $p(x)$.
\end{theorem}

\begin{proof}
	We need to show that the \radup\ defined by the substitution satisfies the \sgdc\ because then \cref{th:hmc_convergence_non-compact} is immediately applicable. For this we show that the update using the substitution reduces to the case covered by \cref{th:exp_pot}, then the \sgdc\ follows.
	
	First, note that a substitution as in \cref{def:eff_pot} is always possible on a \radman\ because any point $x\in X$ can be written as the tuple $x=(r,\theta)$ by definition. Since the substitution $f$ is required to be a diffeomorphism, it is in particular bijective. Thus, for every $r\in[0,\infty)$ there exists a $z\in\mathds{R}$ with $r=f(z)$ and, in consequence, for every $x\in X$ there is a $z\in \mathds{R}$ so that $x=(f(z),\theta)$.
	
	Let us consider the target probability density $p(x)$ under the integral and perform the substitution
	\begin{align}
		\int_X\md x\, p(x) &= \int_\Omega\md \theta\int_0^\infty \md r \left|\det\left(Dx\right)(r,\theta)\right|p(r,\theta)\\
		&= \int_\Omega\md \theta\int_{-\infty}^{\infty} \md z \left|\det\left(Dx\right)(z,\theta)\right|p(f(z),\theta)\\
		&\propto \int_\Omega\md \theta\int_{-\infty}^{\infty} \md z\, \eto{-V(f(z),\theta) + \ln\left|\det J(z,\theta)\right|}\,.
	\end{align}
	Thus, after the substitution, the probability density is defined by the \effpot\ $V_\text{eff}(z,\theta)$ as in equation~\eqref{eq:def_eff_pot}.
	Condition~\eqref{eq:subst_condition} guarantees the applicability of \cref{th:exp_pot} and thus satisfaction of the \sgdc\ for $V_\text{eff}(z,\theta)$ for every fixed $\theta$.
	
	In the Euclidean case, the Jacobi determinant for spherical coordinates can be explicitly incorporated, yielding
	\begin{align}
		\int_X\md x\, p(x) &= \int_\Omega\md \theta\int_0^\infty \md r\, r^{d-1}p(r,\theta)\\
		&= \int_\Omega\md \theta\int_{-\infty}^{\infty} \md z\, f(z)^{d-1} \left|f'(z)\right|p(f(z),\theta)
	\end{align}
	and thus $V_\text{eff}(z,\theta)=V(f(z),\theta) - (d-1)\ln f(z) - \ln\left| f'(z)\right|$, consistent with the definition~\eqref{eq:def_eucl_eff_pot}.
	
	Since the sub-manifold of angles $\Omega$ is compact and all considered functions are continuous, the $\theta$-dependent coefficients $\alpha$ and $K$ in the \sgdc\ as well as $R$ from \cref{th:exp_pot} reach their maxima with $\max_\theta\alpha \eqqcolon\alpha_0 < 1$, $\max_\theta K \eqqcolon K_0<\infty$, and $\max_\theta R \eqqcolon R_0<\infty$. Choose the compact set $X_0 = \left\{x: ||x|| \le R_0\right\}$ for \cref{th:hmc_convergence_non-compact}, then in $X\setminus X_0$ the \sgdc\ is satisfied everywhere, i.e.\ for all angles $\theta$, with the constants $\alpha_0$ and $K_0$. (This last part, that the global $\theta$-dependent satisfaction of the \sgdc\ implies uniform satisfaction of the \sgdc, has been shown in Ref.~\cite{original_radial_update} before.)
\end{proof}

\begin{corollary}[(log-)normal \radup]\label{th:updates_in_practice}
	On a non-compact finite-dimensional Euclidean manifold $X$ any algorithm with the stationary probability distribution $p(x)\propto\eto{-V(x)}$ that satisfies the \cdc\ and the \wgdc\ is guaranteed to converge exponentially to $p(x)$ if it is combined with one of the following \radup s generated by a normal distribution $\gamma \sim \mathcal{N}(0,\sigma^2)$, $\sigma>0$, depending on the potential $V(x)$ (see tab.~\ref{tab:landau_symbols}):
	\begin{itemize}
		\item exponential potential $V(r, \theta) = c\eto{a r+o(r)}$:
		\begin{alignat}{4}
			&r &&\mapsto r+\gamma\,,&&\quad p_\text{acc} &&= \eto{-\Delta V + (d-1) \ln(1+\gamma/r)}
		\end{alignat}
		\item polynomial potential $V(r,\theta) = c r^{a + o(1)}$:
		\begin{alignat}{4}
			&r&&\mapsto r\eto{\gamma}\,,&&\quad p_\text{acc} &&= \eto{-\Delta V + d\gamma}\label{eq:subst_poly_pot}
		\end{alignat}
		\item logarithmic potential $V(r,\theta) = c \ln(r)\eto{o(\log\log r)}$:
		\begin{alignat}{4}
			&r&&\mapsto r^{\eto{\gamma}}\,,&&\quad p_\text{acc} &&= \eto{-\Delta V + d\ln{r}(\eto{\gamma}-1)+\gamma}
		\end{alignat}
	\end{itemize}
	Here $p_\text{acc}$ denotes the acceptance probability in the Metropolis-Hastings step of the \radup\ and $\Delta V$ is the respective change of the potential by the update.
\end{corollary}

\begin{remark}
	Figure~\ref{fig:histograms} shows how \cref{th:updates_in_practice,th:dbl_exp_subst} can be used for the efficient sampling of (heavy-tailed) distributions.
	
	While useful in practice, \cref{th:updates_in_practice} is a much weaker statement than the general \cref{th:dbl_exp_subst}. For instance, a potential of the form $V(r)=d\ln r + 2\ln\ln r$ requires a triple-exponential substitution, as can easily be checked with \cref{th:dbl_exp_subst}, while a simple update similar to those in \cref{th:updates_in_practice} appears unfeasible. Moreover, the updates in \cref{th:updates_in_practice} only consider asymptotics. Based on \cref{th:dbl_exp_subst}, it is easy to construct substitutions that allow to sample the radius $r$ reliably in the entire allowed region $[0,\infty)$. From \cref{th:gauss_update} we know that the update with a normally distributed random number satisfies the \cdc, so no additional \cdc-satisfying algorithm like the HMC is required. Double-exponential formulae useful here can be found in Ref.~\cite{dbl_exp_trafo}. A good choice for polynomial potentials is $r=\exp(z-\eto{-z})$ and for logarithmic potentials $r=\exp(\sinh z)$.
	
	Another advantage of using \cref{th:dbl_exp_subst} is that no tedious calculations of the proposal probabilities in the \mhacc\ are required. The update of the substituted variable is symmetric by construction and the remaining contributions are taken care of by the \effpot\ $V_\text{eff}$. A generalised version of Monte Carlo updates by substitution is discussed in Ref.~\cite{Dutta_2014}.
	
	Of course, the \symker\ for the choice of $\gamma$ in the \radup\ does not have to be a normal distribution either. This choice has been motivated by Ref.~\cite{original_radial_update} where discrete updates $\gamma=\pm\epsilon$ with equal probability have been proposed. In the limit of $\epsilon\rightarrow0$ and infinitely many updates, the random walk of $\gamma$ approaches a normal distribution. The normal distribution appears to perform very well in all the examples we have considered so far, but in principle other distributions might be as good or even better.
\end{remark}

\subsection{Generalisation of convergence behaviour}\label{sec:general_conv}
	For completeness sake, let us generalise \cref{th:dbl_exp_subst} by considering a \radup\ that does not necessarily satisfy the \sgdc. It turns out that, apart from some pathological cases, exponential convergence is guaranteed for practically arbitrary update schemes that satisfy the \cdc\ on \radman s. However, this should not be treated as a justification to use inferior update schemes since they might reduce the speed of convergence by many orders of magnitude (recall figs.~\ref{fig:histograms} and~\ref{fig:ts_hist_lin}). In fact, a \radup\ as in \cref{th:dbl_exp_subst} (or \cref{th:updates_in_practice}) should be used whenever possible. This generalisation is foremost a mathematical curiosity. Its only additional use is in case data has already been generated without an appropriate \radup, e.g.\ using a pure HMC with a polynomial potential. Then \cref{th:classification} (or \cref{th:hmc_conv}) shows that the data does not have to be discarded, provided that thermalisation has been checked carefully.
	
	All the auxiliary definitions, lemmas and proofs leading up to \cref{th:classification} have been moved to \cref{sec:slow_conv}.
	
	\begin{theorem}[Classification of convergence types]\label{th:classification}
		On a \radman\ $X$ combine any algorithm with the stationary probability distribution $p(x)\propto\eto{-V(x)}$ that satisfies the \cdc\ and the
		$\begin{Bmatrix}
			\text{\wgdc} \\ \text{\wadc} \\ \text{\wadc}
		\end{Bmatrix}$
		for the potential $V$ on $X$ with a \radup\ generated by an arbitrary \nontriv\ \symker\ $\mathcal{P}$. Then the combined algorithm converges exponentially to $p(x)$ and it \approach es the region of exponential convergence at least
		$\begin{Bmatrix}
			\text{exponentially} \\ \text{polynomially} \\ \text{diffusively}
		\end{Bmatrix}$
		if the potential $V$ fulfils
		$\begin{Bmatrix}
			\log V(r,\theta)\in\Omega(r)\\
			V(r,\theta)\in\Omega(r)\\
			\eto{-V(r,\theta)} \text{ is normalisable}
		\end{Bmatrix}$
		in spherical coordinates $x=(r,\theta)\in X$.
		Super-exponential \approach\ or convergence cannot be achieved by a \symker.
	\end{theorem}
	
	\begin{corollary}[General HMC convergence]\label{th:hmc_conv}
		The HMC algorithm (without augmentation, e.g.\ by radial updates) on a \radman\ $X$ converges exponentially to the target probability distribution $p(x)\propto\eto{-V(x)}$ and it \approach es the region of exponential convergence at least
		$\begin{Bmatrix}
			\text{exponentially} \\ \text{polynomially} \\ \text{diffusively}
		\end{Bmatrix}$
		if the smooth potential $V$ fulfils
		$\begin{Bmatrix}
			\log V(r,\theta)\in\Omega(r)\\
			V(r,\theta)\in\Omega(r)\\
			\eto{-V(r,\theta)} \text{ is normalisable}
		\end{Bmatrix}$.
	\end{corollary}
	
\begin{figure}[th]
		\centering
		\resizebox{0.98\textwidth}{!}{{\large%
\begingroup
  \inputencoding{latin1}%
  \makeatletter
  \providecommand\color[2][]{%
    \GenericError{(gnuplot) \space\space\space\@spaces}{%
      Package color not loaded in conjunction with
      terminal option `colourtext'%
    }{See the gnuplot documentation for explanation.%
    }{Either use 'blacktext' in gnuplot or load the package
      color.sty in LaTeX.}%
    \renewcommand\color[2][]{}%
  }%
  \providecommand\includegraphics[2][]{%
    \GenericError{(gnuplot) \space\space\space\@spaces}{%
      Package graphicx or graphics not loaded%
    }{See the gnuplot documentation for explanation.%
    }{The gnuplot epslatex terminal needs graphicx.sty or graphics.sty.}%
    \renewcommand\includegraphics[2][]{}%
  }%
  \providecommand\rotatebox[2]{#2}%
  \@ifundefined{ifGPcolor}{%
    \newif\ifGPcolor
    \GPcolortrue
  }{}%
  \@ifundefined{ifGPblacktext}{%
    \newif\ifGPblacktext
    \GPblacktexttrue
  }{}%
  \let\gplgaddtomacro\g@addto@macro
  \gdef\gplbacktext{}%
  \gdef\gplfronttext{}%
  \makeatother
  \ifGPblacktext
    \def\colorrgb#1{}%
    \def\colorgray#1{}%
  \else
    \ifGPcolor
      \def\colorrgb#1{\color[rgb]{#1}}%
      \def\colorgray#1{\color[gray]{#1}}%
      \expandafter\def\csname LTw\endcsname{\color{white}}%
      \expandafter\def\csname LTb\endcsname{\color{black}}%
      \expandafter\def\csname LTa\endcsname{\color{black}}%
      \expandafter\def\csname LT0\endcsname{\color[rgb]{1,0,0}}%
      \expandafter\def\csname LT1\endcsname{\color[rgb]{0,1,0}}%
      \expandafter\def\csname LT2\endcsname{\color[rgb]{0,0,1}}%
      \expandafter\def\csname LT3\endcsname{\color[rgb]{1,0,1}}%
      \expandafter\def\csname LT4\endcsname{\color[rgb]{0,1,1}}%
      \expandafter\def\csname LT5\endcsname{\color[rgb]{1,1,0}}%
      \expandafter\def\csname LT6\endcsname{\color[rgb]{0,0,0}}%
      \expandafter\def\csname LT7\endcsname{\color[rgb]{1,0.3,0}}%
      \expandafter\def\csname LT8\endcsname{\color[rgb]{0.5,0.5,0.5}}%
    \else
      \def\colorrgb#1{\color{black}}%
      \def\colorgray#1{\color[gray]{#1}}%
      \expandafter\def\csname LTw\endcsname{\color{white}}%
      \expandafter\def\csname LTb\endcsname{\color{black}}%
      \expandafter\def\csname LTa\endcsname{\color{black}}%
      \expandafter\def\csname LT0\endcsname{\color{black}}%
      \expandafter\def\csname LT1\endcsname{\color{black}}%
      \expandafter\def\csname LT2\endcsname{\color{black}}%
      \expandafter\def\csname LT3\endcsname{\color{black}}%
      \expandafter\def\csname LT4\endcsname{\color{black}}%
      \expandafter\def\csname LT5\endcsname{\color{black}}%
      \expandafter\def\csname LT6\endcsname{\color{black}}%
      \expandafter\def\csname LT7\endcsname{\color{black}}%
      \expandafter\def\csname LT8\endcsname{\color{black}}%
    \fi
  \fi
    \setlength{\unitlength}{0.0500bp}%
    \ifx\gptboxheight\undefined%
      \newlength{\gptboxheight}%
      \newlength{\gptboxwidth}%
      \newsavebox{\gptboxtext}%
    \fi%
    \setlength{\fboxrule}{0.5pt}%
    \setlength{\fboxsep}{1pt}%
    \definecolor{tbcol}{rgb}{1,1,1}%
\begin{picture}(7200.00,5040.00)%
    \gplgaddtomacro\gplbacktext{%
      \csname LTb\endcsname%
      \put(946,704){\makebox(0,0)[r]{\strut{}$10^{0}$}}%
      \csname LTb\endcsname%
      \put(946,1177){\makebox(0,0)[r]{\strut{}$10^{10}$}}%
      \csname LTb\endcsname%
      \put(946,1650){\makebox(0,0)[r]{\strut{}$10^{20}$}}%
      \csname LTb\endcsname%
      \put(946,2123){\makebox(0,0)[r]{\strut{}$10^{30}$}}%
      \csname LTb\endcsname%
      \put(946,2596){\makebox(0,0)[r]{\strut{}$10^{40}$}}%
      \csname LTb\endcsname%
      \put(946,3069){\makebox(0,0)[r]{\strut{}$10^{50}$}}%
      \csname LTb\endcsname%
      \put(946,3542){\makebox(0,0)[r]{\strut{}$10^{60}$}}%
      \csname LTb\endcsname%
      \put(946,4015){\makebox(0,0)[r]{\strut{}$10^{70}$}}%
      \csname LTb\endcsname%
      \put(946,4488){\makebox(0,0)[r]{\strut{}$10^{80}$}}%
      \csname LTb\endcsname%
      \put(1078,484){\makebox(0,0){\strut{}$0$}}%
      \csname LTb\endcsname%
      \put(2032,484){\makebox(0,0){\strut{}$100$}}%
      \csname LTb\endcsname%
      \put(2986,484){\makebox(0,0){\strut{}$200$}}%
      \csname LTb\endcsname%
      \put(3941,484){\makebox(0,0){\strut{}$300$}}%
      \csname LTb\endcsname%
      \put(4895,484){\makebox(0,0){\strut{}$400$}}%
      \csname LTb\endcsname%
      \put(5849,484){\makebox(0,0){\strut{}$500$}}%
      \csname LTb\endcsname%
      \put(6803,484){\makebox(0,0){\strut{}$600$}}%
    }%
    \gplgaddtomacro\gplfronttext{%
      \csname LTb\endcsname%
      \put(209,2761){\rotatebox{-270.00}{\makebox(0,0){\strut{}$V(z)=\cosh(z)$}}}%
      \put(3940,154){\makebox(0,0){\strut{}$t$}}%
    }%
    \gplbacktext
    \put(0,0){\includegraphics[width={360.00bp},height={252.00bp}]{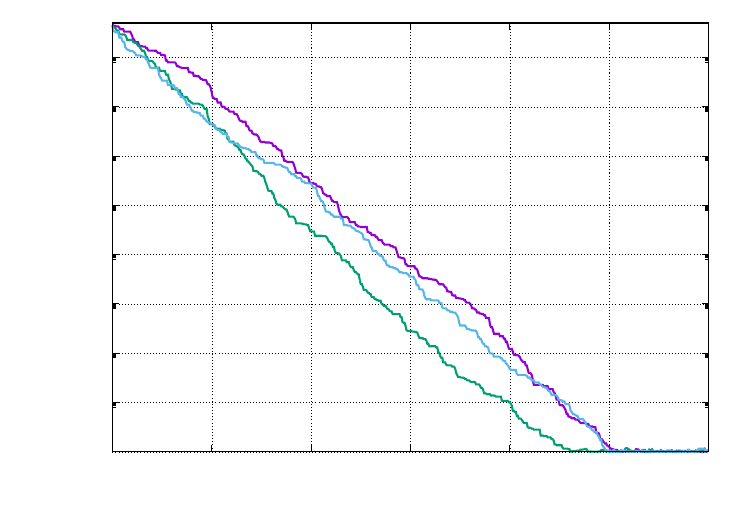}}%
    \gplfronttext
  \end{picture}%
\endgroup
\begingroup
  \inputencoding{latin1}%
  \makeatletter
  \providecommand\color[2][]{%
    \GenericError{(gnuplot) \space\space\space\@spaces}{%
      Package color not loaded in conjunction with
      terminal option `colourtext'%
    }{See the gnuplot documentation for explanation.%
    }{Either use 'blacktext' in gnuplot or load the package
      color.sty in LaTeX.}%
    \renewcommand\color[2][]{}%
  }%
  \providecommand\includegraphics[2][]{%
    \GenericError{(gnuplot) \space\space\space\@spaces}{%
      Package graphicx or graphics not loaded%
    }{See the gnuplot documentation for explanation.%
    }{The gnuplot epslatex terminal needs graphicx.sty or graphics.sty.}%
    \renewcommand\includegraphics[2][]{}%
  }%
  \providecommand\rotatebox[2]{#2}%
  \@ifundefined{ifGPcolor}{%
    \newif\ifGPcolor
    \GPcolortrue
  }{}%
  \@ifundefined{ifGPblacktext}{%
    \newif\ifGPblacktext
    \GPblacktexttrue
  }{}%
  \let\gplgaddtomacro\g@addto@macro
  \gdef\gplbacktext{}%
  \gdef\gplfronttext{}%
  \makeatother
  \ifGPblacktext
    \def\colorrgb#1{}%
    \def\colorgray#1{}%
  \else
    \ifGPcolor
      \def\colorrgb#1{\color[rgb]{#1}}%
      \def\colorgray#1{\color[gray]{#1}}%
      \expandafter\def\csname LTw\endcsname{\color{white}}%
      \expandafter\def\csname LTb\endcsname{\color{black}}%
      \expandafter\def\csname LTa\endcsname{\color{black}}%
      \expandafter\def\csname LT0\endcsname{\color[rgb]{1,0,0}}%
      \expandafter\def\csname LT1\endcsname{\color[rgb]{0,1,0}}%
      \expandafter\def\csname LT2\endcsname{\color[rgb]{0,0,1}}%
      \expandafter\def\csname LT3\endcsname{\color[rgb]{1,0,1}}%
      \expandafter\def\csname LT4\endcsname{\color[rgb]{0,1,1}}%
      \expandafter\def\csname LT5\endcsname{\color[rgb]{1,1,0}}%
      \expandafter\def\csname LT6\endcsname{\color[rgb]{0,0,0}}%
      \expandafter\def\csname LT7\endcsname{\color[rgb]{1,0.3,0}}%
      \expandafter\def\csname LT8\endcsname{\color[rgb]{0.5,0.5,0.5}}%
    \else
      \def\colorrgb#1{\color{black}}%
      \def\colorgray#1{\color[gray]{#1}}%
      \expandafter\def\csname LTw\endcsname{\color{white}}%
      \expandafter\def\csname LTb\endcsname{\color{black}}%
      \expandafter\def\csname LTa\endcsname{\color{black}}%
      \expandafter\def\csname LT0\endcsname{\color{black}}%
      \expandafter\def\csname LT1\endcsname{\color{black}}%
      \expandafter\def\csname LT2\endcsname{\color{black}}%
      \expandafter\def\csname LT3\endcsname{\color{black}}%
      \expandafter\def\csname LT4\endcsname{\color{black}}%
      \expandafter\def\csname LT5\endcsname{\color{black}}%
      \expandafter\def\csname LT6\endcsname{\color{black}}%
      \expandafter\def\csname LT7\endcsname{\color{black}}%
      \expandafter\def\csname LT8\endcsname{\color{black}}%
    \fi
  \fi
    \setlength{\unitlength}{0.0500bp}%
    \ifx\gptboxheight\undefined%
      \newlength{\gptboxheight}%
      \newlength{\gptboxwidth}%
      \newsavebox{\gptboxtext}%
    \fi%
    \setlength{\fboxrule}{0.5pt}%
    \setlength{\fboxsep}{1pt}%
    \definecolor{tbcol}{rgb}{1,1,1}%
\begin{picture}(7200.00,5040.00)%
    \gplgaddtomacro\gplbacktext{%
      \csname LTb\endcsname%
      \put(814,704){\makebox(0,0)[r]{\strut{}$0$}}%
      \csname LTb\endcsname%
      \put(814,1527){\makebox(0,0)[r]{\strut{}$50$}}%
      \csname LTb\endcsname%
      \put(814,2350){\makebox(0,0)[r]{\strut{}$100$}}%
      \csname LTb\endcsname%
      \put(814,3173){\makebox(0,0)[r]{\strut{}$150$}}%
      \csname LTb\endcsname%
      \put(814,3996){\makebox(0,0)[r]{\strut{}$200$}}%
      \csname LTb\endcsname%
      \put(814,4819){\makebox(0,0)[r]{\strut{}$250$}}%
      \csname LTb\endcsname%
      \put(946,484){\makebox(0,0){\strut{}$0$}}%
      \csname LTb\endcsname%
      \put(2117,484){\makebox(0,0){\strut{}$200$}}%
      \csname LTb\endcsname%
      \put(3289,484){\makebox(0,0){\strut{}$400$}}%
      \csname LTb\endcsname%
      \put(4460,484){\makebox(0,0){\strut{}$600$}}%
      \csname LTb\endcsname%
      \put(5632,484){\makebox(0,0){\strut{}$800$}}%
      \csname LTb\endcsname%
      \put(6803,484){\makebox(0,0){\strut{}$1000$}}%
    }%
    \gplgaddtomacro\gplfronttext{%
      \csname LTb\endcsname%
      \put(209,2761){\rotatebox{-270.00}{\makebox(0,0){\strut{}$V(z)=|z|$}}}%
      \put(3874,154){\makebox(0,0){\strut{}$t$}}%
    }%
    \gplbacktext
    \put(0,0){\includegraphics[width={360.00bp},height={252.00bp}]{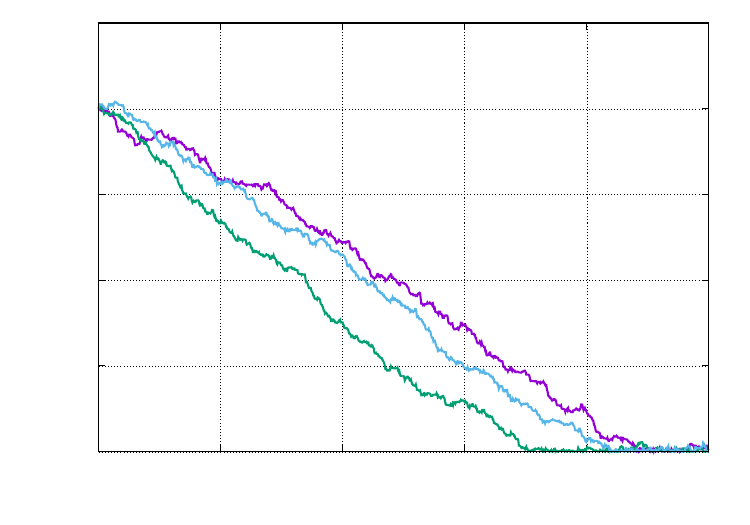}}%
    \gplfronttext
  \end{picture}%
\endgroup
}}
		\resizebox{0.98\textwidth}{!}{{\large%
\begingroup
  \inputencoding{latin1}%
  \makeatletter
  \providecommand\color[2][]{%
    \GenericError{(gnuplot) \space\space\space\@spaces}{%
      Package color not loaded in conjunction with
      terminal option `colourtext'%
    }{See the gnuplot documentation for explanation.%
    }{Either use 'blacktext' in gnuplot or load the package
      color.sty in LaTeX.}%
    \renewcommand\color[2][]{}%
  }%
  \providecommand\includegraphics[2][]{%
    \GenericError{(gnuplot) \space\space\space\@spaces}{%
      Package graphicx or graphics not loaded%
    }{See the gnuplot documentation for explanation.%
    }{The gnuplot epslatex terminal needs graphicx.sty or graphics.sty.}%
    \renewcommand\includegraphics[2][]{}%
  }%
  \providecommand\rotatebox[2]{#2}%
  \@ifundefined{ifGPcolor}{%
    \newif\ifGPcolor
    \GPcolortrue
  }{}%
  \@ifundefined{ifGPblacktext}{%
    \newif\ifGPblacktext
    \GPblacktexttrue
  }{}%
  \let\gplgaddtomacro\g@addto@macro
  \gdef\gplbacktext{}%
  \gdef\gplfronttext{}%
  \makeatother
  \ifGPblacktext
    \def\colorrgb#1{}%
    \def\colorgray#1{}%
  \else
    \ifGPcolor
      \def\colorrgb#1{\color[rgb]{#1}}%
      \def\colorgray#1{\color[gray]{#1}}%
      \expandafter\def\csname LTw\endcsname{\color{white}}%
      \expandafter\def\csname LTb\endcsname{\color{black}}%
      \expandafter\def\csname LTa\endcsname{\color{black}}%
      \expandafter\def\csname LT0\endcsname{\color[rgb]{1,0,0}}%
      \expandafter\def\csname LT1\endcsname{\color[rgb]{0,1,0}}%
      \expandafter\def\csname LT2\endcsname{\color[rgb]{0,0,1}}%
      \expandafter\def\csname LT3\endcsname{\color[rgb]{1,0,1}}%
      \expandafter\def\csname LT4\endcsname{\color[rgb]{0,1,1}}%
      \expandafter\def\csname LT5\endcsname{\color[rgb]{1,1,0}}%
      \expandafter\def\csname LT6\endcsname{\color[rgb]{0,0,0}}%
      \expandafter\def\csname LT7\endcsname{\color[rgb]{1,0.3,0}}%
      \expandafter\def\csname LT8\endcsname{\color[rgb]{0.5,0.5,0.5}}%
    \else
      \def\colorrgb#1{\color{black}}%
      \def\colorgray#1{\color[gray]{#1}}%
      \expandafter\def\csname LTw\endcsname{\color{white}}%
      \expandafter\def\csname LTb\endcsname{\color{black}}%
      \expandafter\def\csname LTa\endcsname{\color{black}}%
      \expandafter\def\csname LT0\endcsname{\color{black}}%
      \expandafter\def\csname LT1\endcsname{\color{black}}%
      \expandafter\def\csname LT2\endcsname{\color{black}}%
      \expandafter\def\csname LT3\endcsname{\color{black}}%
      \expandafter\def\csname LT4\endcsname{\color{black}}%
      \expandafter\def\csname LT5\endcsname{\color{black}}%
      \expandafter\def\csname LT6\endcsname{\color{black}}%
      \expandafter\def\csname LT7\endcsname{\color{black}}%
      \expandafter\def\csname LT8\endcsname{\color{black}}%
    \fi
  \fi
    \setlength{\unitlength}{0.0500bp}%
    \ifx\gptboxheight\undefined%
      \newlength{\gptboxheight}%
      \newlength{\gptboxwidth}%
      \newsavebox{\gptboxtext}%
    \fi%
    \setlength{\fboxrule}{0.5pt}%
    \setlength{\fboxsep}{1pt}%
    \definecolor{tbcol}{rgb}{1,1,1}%
\begin{picture}(7200.00,5040.00)%
    \gplgaddtomacro\gplbacktext{%
      \csname LTb\endcsname%
      \put(682,704){\makebox(0,0)[r]{\strut{}$0$}}%
      \csname LTb\endcsname%
      \put(682,1161){\makebox(0,0)[r]{\strut{}$2$}}%
      \csname LTb\endcsname%
      \put(682,1618){\makebox(0,0)[r]{\strut{}$4$}}%
      \csname LTb\endcsname%
      \put(682,2076){\makebox(0,0)[r]{\strut{}$6$}}%
      \csname LTb\endcsname%
      \put(682,2533){\makebox(0,0)[r]{\strut{}$8$}}%
      \csname LTb\endcsname%
      \put(682,2990){\makebox(0,0)[r]{\strut{}$10$}}%
      \csname LTb\endcsname%
      \put(682,3447){\makebox(0,0)[r]{\strut{}$12$}}%
      \csname LTb\endcsname%
      \put(682,3905){\makebox(0,0)[r]{\strut{}$14$}}%
      \csname LTb\endcsname%
      \put(682,4362){\makebox(0,0)[r]{\strut{}$16$}}%
      \csname LTb\endcsname%
      \put(682,4819){\makebox(0,0)[r]{\strut{}$18$}}%
      \csname LTb\endcsname%
      \put(814,484){\makebox(0,0){\strut{}$0$}}%
      \csname LTb\endcsname%
      \put(1613,484){\makebox(0,0){\strut{}$2000$}}%
      \csname LTb\endcsname%
      \put(2411,484){\makebox(0,0){\strut{}$4000$}}%
      \csname LTb\endcsname%
      \put(3210,484){\makebox(0,0){\strut{}$6000$}}%
      \csname LTb\endcsname%
      \put(4008,484){\makebox(0,0){\strut{}$8000$}}%
      \csname LTb\endcsname%
      \put(4807,484){\makebox(0,0){\strut{}$10000$}}%
      \csname LTb\endcsname%
      \put(5605,484){\makebox(0,0){\strut{}$12000$}}%
      \csname LTb\endcsname%
      \put(6404,484){\makebox(0,0){\strut{}$14000$}}%
    }%
    \gplgaddtomacro\gplfronttext{%
      \csname LTb\endcsname%
      \put(209,2761){\rotatebox{-270.00}{\makebox(0,0){\strut{}$V(z)=\sqrt{|z|}$}}}%
      \put(3808,154){\makebox(0,0){\strut{}$t$}}%
    }%
    \gplbacktext
    \put(0,0){\includegraphics[width={360.00bp},height={252.00bp}]{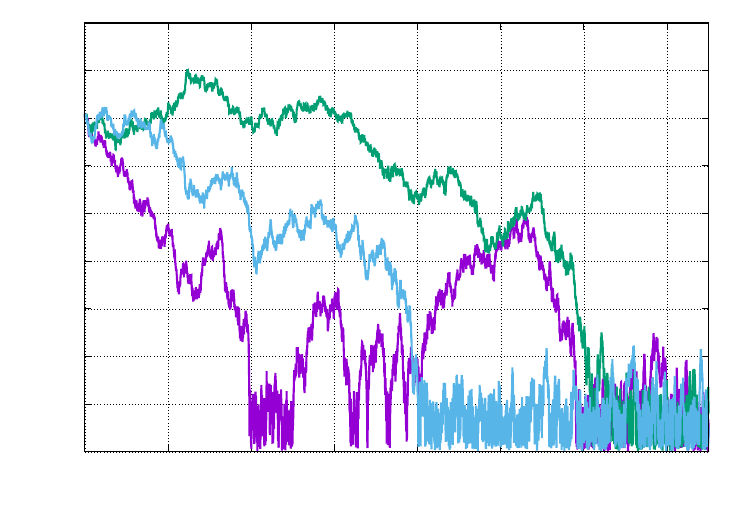}}%
    \gplfronttext
  \end{picture}%
\endgroup
\begingroup
  \inputencoding{latin1}%
  \makeatletter
  \providecommand\color[2][]{%
    \GenericError{(gnuplot) \space\space\space\@spaces}{%
      Package color not loaded in conjunction with
      terminal option `colourtext'%
    }{See the gnuplot documentation for explanation.%
    }{Either use 'blacktext' in gnuplot or load the package
      color.sty in LaTeX.}%
    \renewcommand\color[2][]{}%
  }%
  \providecommand\includegraphics[2][]{%
    \GenericError{(gnuplot) \space\space\space\@spaces}{%
      Package graphicx or graphics not loaded%
    }{See the gnuplot documentation for explanation.%
    }{The gnuplot epslatex terminal needs graphicx.sty or graphics.sty.}%
    \renewcommand\includegraphics[2][]{}%
  }%
  \providecommand\rotatebox[2]{#2}%
  \@ifundefined{ifGPcolor}{%
    \newif\ifGPcolor
    \GPcolortrue
  }{}%
  \@ifundefined{ifGPblacktext}{%
    \newif\ifGPblacktext
    \GPblacktexttrue
  }{}%
  \let\gplgaddtomacro\g@addto@macro
  \gdef\gplbacktext{}%
  \gdef\gplfronttext{}%
  \makeatother
  \ifGPblacktext
    \def\colorrgb#1{}%
    \def\colorgray#1{}%
  \else
    \ifGPcolor
      \def\colorrgb#1{\color[rgb]{#1}}%
      \def\colorgray#1{\color[gray]{#1}}%
      \expandafter\def\csname LTw\endcsname{\color{white}}%
      \expandafter\def\csname LTb\endcsname{\color{black}}%
      \expandafter\def\csname LTa\endcsname{\color{black}}%
      \expandafter\def\csname LT0\endcsname{\color[rgb]{1,0,0}}%
      \expandafter\def\csname LT1\endcsname{\color[rgb]{0,1,0}}%
      \expandafter\def\csname LT2\endcsname{\color[rgb]{0,0,1}}%
      \expandafter\def\csname LT3\endcsname{\color[rgb]{1,0,1}}%
      \expandafter\def\csname LT4\endcsname{\color[rgb]{0,1,1}}%
      \expandafter\def\csname LT5\endcsname{\color[rgb]{1,1,0}}%
      \expandafter\def\csname LT6\endcsname{\color[rgb]{0,0,0}}%
      \expandafter\def\csname LT7\endcsname{\color[rgb]{1,0.3,0}}%
      \expandafter\def\csname LT8\endcsname{\color[rgb]{0.5,0.5,0.5}}%
    \else
      \def\colorrgb#1{\color{black}}%
      \def\colorgray#1{\color[gray]{#1}}%
      \expandafter\def\csname LTw\endcsname{\color{white}}%
      \expandafter\def\csname LTb\endcsname{\color{black}}%
      \expandafter\def\csname LTa\endcsname{\color{black}}%
      \expandafter\def\csname LT0\endcsname{\color{black}}%
      \expandafter\def\csname LT1\endcsname{\color{black}}%
      \expandafter\def\csname LT2\endcsname{\color{black}}%
      \expandafter\def\csname LT3\endcsname{\color{black}}%
      \expandafter\def\csname LT4\endcsname{\color{black}}%
      \expandafter\def\csname LT5\endcsname{\color{black}}%
      \expandafter\def\csname LT6\endcsname{\color{black}}%
      \expandafter\def\csname LT7\endcsname{\color{black}}%
      \expandafter\def\csname LT8\endcsname{\color{black}}%
    \fi
  \fi
    \setlength{\unitlength}{0.0500bp}%
    \ifx\gptboxheight\undefined%
      \newlength{\gptboxheight}%
      \newlength{\gptboxwidth}%
      \newsavebox{\gptboxtext}%
    \fi%
    \setlength{\fboxrule}{0.5pt}%
    \setlength{\fboxsep}{1pt}%
    \definecolor{tbcol}{rgb}{1,1,1}%
\begin{picture}(7200.00,5040.00)%
    \gplgaddtomacro\gplbacktext{%
      \csname LTb\endcsname%
      \put(682,704){\makebox(0,0)[r]{\strut{}$0$}}%
      \csname LTb\endcsname%
      \put(682,1390){\makebox(0,0)[r]{\strut{}$2$}}%
      \csname LTb\endcsname%
      \put(682,2076){\makebox(0,0)[r]{\strut{}$4$}}%
      \csname LTb\endcsname%
      \put(682,2762){\makebox(0,0)[r]{\strut{}$6$}}%
      \csname LTb\endcsname%
      \put(682,3447){\makebox(0,0)[r]{\strut{}$8$}}%
      \csname LTb\endcsname%
      \put(682,4133){\makebox(0,0)[r]{\strut{}$10$}}%
      \csname LTb\endcsname%
      \put(682,4819){\makebox(0,0)[r]{\strut{}$12$}}%
      \csname LTb\endcsname%
      \put(814,484){\makebox(0,0){\strut{}$0$}}%
      \csname LTb\endcsname%
      \put(1735,484){\makebox(0,0){\strut{}$10000$}}%
      \csname LTb\endcsname%
      \put(2657,484){\makebox(0,0){\strut{}$20000$}}%
      \csname LTb\endcsname%
      \put(3578,484){\makebox(0,0){\strut{}$30000$}}%
      \csname LTb\endcsname%
      \put(4500,484){\makebox(0,0){\strut{}$40000$}}%
      \csname LTb\endcsname%
      \put(5421,484){\makebox(0,0){\strut{}$50000$}}%
      \csname LTb\endcsname%
      \put(6342,484){\makebox(0,0){\strut{}$60000$}}%
    }%
    \gplgaddtomacro\gplfronttext{%
      \csname LTb\endcsname%
      \put(209,2761){\rotatebox{-270.00}{\makebox(0,0){\strut{}$V(z)=\ln\left(1+z^2\right)$}}}%
      \put(3808,154){\makebox(0,0){\strut{}$t$}}%
    }%
    \gplbacktext
    \put(0,0){\includegraphics[width={360.00bp},height={252.00bp}]{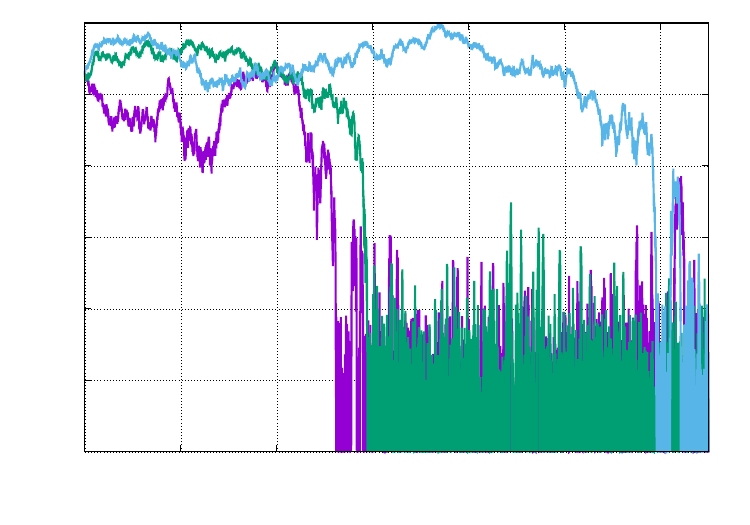}}%
    \gplfronttext
  \end{picture}%
\endgroup
}}
		\caption{Convergence of the Lyapunov function~\cite{Lyapounov1907,Meyn_Tweedie_1992,meyn1993markov} (equivalent to the potential) for different potentials in $d=1$ dimension. In all cases the update $z\rightarrow z+\gamma$ with $\gamma\sim \mathcal{N}(0,1)$ has been used. The different time series in every panel correspond to different random number seeds. The potentials
			$V(z)=\begin{Bmatrix}
				\cosh(z) \\ |z| \\ \sqrt{|z|} \text{ and } \ln\left(1+z^2\right)
			\end{Bmatrix}$ in the
			$\begin{Bmatrix}
				\text{top left} \\ \text{top right} \\ \text{bottom}
			\end{Bmatrix}$ panels grow
			$\begin{Bmatrix}
				\text{exponentially} \\ \text{linearly} \\ \text{sub-linearly}
			\end{Bmatrix}$ and thus satisfy the
			$\begin{Bmatrix}
				\text{\sgdc} \\ \text{\sadc} \\ \text{\ladc}
			\end{Bmatrix}$ leading to an
			$\begin{Bmatrix}
				\text{exponential} \\ \text{linear} \\ \text{diffusive}
			\end{Bmatrix}$ \approach\ of the exponential convergence region according to \cref{th:classification}.}\label{fig:approaches}
	\end{figure}

	\Cref{fig:approaches} visualises the different types of \approach es towards the correct probability distribution. The same normal additive update has been used to sample from distributions induced by different potentials. The distance from the target probability distribution is bounded by the expectation value of the potential~\cite{hairer2008look}, interpreted as a Lyapunov function~\cite{Lyapounov1907,Meyn_Tweedie_1992,meyn1993markov} in this setting. Thus, an ensemble of different realisations of the time series conveys a representative impression of the convergence in probability, mediated by the expected potential.
	
	One can clearly observe that, on the one hand, all simulations eventually converge since all random realisation \approach\ a thermodynamic equilibrium close to the minimum of the potential. On the other hand, this \approach\ is qualitatively different for the four chosen potentials. In the upper two panels a clear drift forces all the time series to decrease at a very similar and constant rate (exponentially in the left panel and linearly in the right one). In contrast, in the lower panels no such drift is apparent and the \approach\ resembles a random walk. Note that the \approach es in the bottom two panels are qualitatively very similar, as expected from \cref{th:classification}, even though the potentials are strikingly different (sub-linear polynomial on the left and logarithmic on the right).
	
\subsection{Parameter tuning}\label{sec:param_tuning}
	After considering the most general case, let us focus on the most common case. In physics, most potentials are typically polynomial. In this case one should use the update $r\mapsto r\eto{\gamma}$, or equivalently, sample the new radius $r'$ from the log-normal distribution $r'\sim \mathrm{Lognormal}(\ln r, \sigma^2)$. Now $\sigma$ is a free parameter and, according to \cref{th:updates_in_practice}, any $\sigma>0$ guarantees a valid update scheme. In practice, however, we are interested in a parameter choice that optimises the sampling efficiency, for instance via minimisation of the \intaut. In general, it is impossible to calculate the exact optimal value of $\sigma$ analytically, however the scaling of $\sigma$ with the dimension $d$ can be derived.
	
	\begin{definition}[Integrated autocorrelation time~\cite{WOLFF2004143}]\label{def:tau_int}
		For a time series of the observable $\mathcal{A}$ with MCMC time $t$ define the integrated autocorrelation time
		\begin{align}
			\tint &\equiv \frac12 + \sum_{t=1}^{\infty} C_\mathcal{A}(t)\,,
		\end{align}
		where $C_\mathcal{A}(t)$ denotes the autocorrelation function of $\mathcal{A}$.
	\end{definition}
	
	The \intaut\ is arguably the most important parameter for the quantification of the sampling efficiency in MCMC simulations. In practice, the sum has to be truncated. A reliable way to do so with minimal statistical and systematic errors is described in Ref.~\cite{WOLFF2004143} and the implementation in~\cite{comp-avg} has been used. \tint\ effectively measures how long it will take the Markov chain to produce an independent configuration. That is, given a precision goal, the required compute time is simply proportional to \tint. The best possible sampling decorrelates successive configurations completely and thus results in the lowest possible value of $\tint=\frac12$.
	
	\begin{estimate}[Maximal decorrelation by additive update]\label{th:max_decorr}
		Let $p(z)$ be a 1-dimensional target probability distribution with mean $\mu$ and variance $\sigma_z^2$ that decays quickly in $|z-\mu|$. The update $z\mapsto z+\gamma$ with $\gamma\sim\mathcal{N}(0,\sigma^2)$ that minimises the expected correlation of $z$ and $z+\gamma$ (including Metropolis accept/reject) uses
		\begin{align}
			\sigma &\simeq \sqrt{2}\sigma_z\,.
		\end{align}
	\end{estimate}
	
	\begin{remark}
		The condition of quick decay is needed to derive the prefactor $\sqrt{2}$ analytically. However, numerical experiments suggest that $\sigma_z < \sigma < 2\sigma_z$ holds universally and the prefactor $\sqrt{2}$ is very close to optimal in most cases.
	\end{remark}
	
	\begin{corollary}\label{th:opt_acc}
		Using the parameters of \cref{th:max_decorr}, the correlation after a single update step reduces to
		\begin{align}
			\mathrm{corr}(z,z+\gamma) &\simeq 1-\frac{1-e^{-1}}{\sqrt{\pi}}\approx\num{0.64}\,,\label{eq:corr_z_gamma_opt}
		\end{align}
		resulting in an \intaut\
		\begin{align}
			\tint &\simeq \frac{\sqrt{\pi}}{1-e^{-1}}-\frac12 \approx \num{2.3}
		\end{align}
		and the acceptance rate is
		\begin{align}
			p_\text{acc} &\gtrsim \frac12\mathrm{erf}(1)\approx\num{0.42}\,.
		\end{align}
	\end{corollary}
	
	\begin{theorem}[Scaling of log-normal $\sigma$ for polynomial potentials]\label{th:scaling_sigma}
		Given the update scheme $r\mapsto r\eto{\gamma}$ with $\gamma \sim \mathcal{N}(0,\sigma^2)$ as in \cref{th:updates_in_practice} for a polynomial potential $V(r,\theta) = c r^{a + o(1)}$, the optimal choice of the standard deviation is
		\begin{align}
			\sigma &= \sqrt{\frac{2}{ad}} + \ordnung{d^{-1}}\,,\label{eq:best_sigma}
		\end{align}
		where the numerical factor $\sqrt{2}$ is based on \cref{th:max_decorr} and can vary slightly depending on the potential.
	\end{theorem}
	
	\begin{remark}
		If in doubt, it is advisable to choose $\sigma$ slightly too large rather than too small. Larger standard deviations merely reduce the acceptance rate linearly while smaller $\sigma$ lead to a diffusive regime in which autocorrelations grow quadratically.
	\end{remark}
	
	\begin{figure}[th]
		\centering
		\resizebox{0.98\textwidth}{!}{{\large%
\begingroup
  \inputencoding{latin1}%
  \makeatletter
  \providecommand\color[2][]{%
    \GenericError{(gnuplot) \space\space\space\@spaces}{%
      Package color not loaded in conjunction with
      terminal option `colourtext'%
    }{See the gnuplot documentation for explanation.%
    }{Either use 'blacktext' in gnuplot or load the package
      color.sty in LaTeX.}%
    \renewcommand\color[2][]{}%
  }%
  \providecommand\includegraphics[2][]{%
    \GenericError{(gnuplot) \space\space\space\@spaces}{%
      Package graphicx or graphics not loaded%
    }{See the gnuplot documentation for explanation.%
    }{The gnuplot epslatex terminal needs graphicx.sty or graphics.sty.}%
    \renewcommand\includegraphics[2][]{}%
  }%
  \providecommand\rotatebox[2]{#2}%
  \@ifundefined{ifGPcolor}{%
    \newif\ifGPcolor
    \GPcolortrue
  }{}%
  \@ifundefined{ifGPblacktext}{%
    \newif\ifGPblacktext
    \GPblacktexttrue
  }{}%
  \let\gplgaddtomacro\g@addto@macro
  \gdef\gplbacktext{}%
  \gdef\gplfronttext{}%
  \makeatother
  \ifGPblacktext
    \def\colorrgb#1{}%
    \def\colorgray#1{}%
  \else
    \ifGPcolor
      \def\colorrgb#1{\color[rgb]{#1}}%
      \def\colorgray#1{\color[gray]{#1}}%
      \expandafter\def\csname LTw\endcsname{\color{white}}%
      \expandafter\def\csname LTb\endcsname{\color{black}}%
      \expandafter\def\csname LTa\endcsname{\color{black}}%
      \expandafter\def\csname LT0\endcsname{\color[rgb]{1,0,0}}%
      \expandafter\def\csname LT1\endcsname{\color[rgb]{0,1,0}}%
      \expandafter\def\csname LT2\endcsname{\color[rgb]{0,0,1}}%
      \expandafter\def\csname LT3\endcsname{\color[rgb]{1,0,1}}%
      \expandafter\def\csname LT4\endcsname{\color[rgb]{0,1,1}}%
      \expandafter\def\csname LT5\endcsname{\color[rgb]{1,1,0}}%
      \expandafter\def\csname LT6\endcsname{\color[rgb]{0,0,0}}%
      \expandafter\def\csname LT7\endcsname{\color[rgb]{1,0.3,0}}%
      \expandafter\def\csname LT8\endcsname{\color[rgb]{0.5,0.5,0.5}}%
    \else
      \def\colorrgb#1{\color{black}}%
      \def\colorgray#1{\color[gray]{#1}}%
      \expandafter\def\csname LTw\endcsname{\color{white}}%
      \expandafter\def\csname LTb\endcsname{\color{black}}%
      \expandafter\def\csname LTa\endcsname{\color{black}}%
      \expandafter\def\csname LT0\endcsname{\color{black}}%
      \expandafter\def\csname LT1\endcsname{\color{black}}%
      \expandafter\def\csname LT2\endcsname{\color{black}}%
      \expandafter\def\csname LT3\endcsname{\color{black}}%
      \expandafter\def\csname LT4\endcsname{\color{black}}%
      \expandafter\def\csname LT5\endcsname{\color{black}}%
      \expandafter\def\csname LT6\endcsname{\color{black}}%
      \expandafter\def\csname LT7\endcsname{\color{black}}%
      \expandafter\def\csname LT8\endcsname{\color{black}}%
    \fi
  \fi
    \setlength{\unitlength}{0.0500bp}%
    \ifx\gptboxheight\undefined%
      \newlength{\gptboxheight}%
      \newlength{\gptboxwidth}%
      \newsavebox{\gptboxtext}%
    \fi%
    \setlength{\fboxrule}{0.5pt}%
    \setlength{\fboxsep}{1pt}%
    \definecolor{tbcol}{rgb}{1,1,1}%
\begin{picture}(7200.00,5040.00)%
    \gplgaddtomacro\gplbacktext{%
      \csname LTb\endcsname%
      \put(946,704){\makebox(0,0)[r]{\strut{}$8$}}%
      \csname LTb\endcsname%
      \put(946,1218){\makebox(0,0)[r]{\strut{}$8.5$}}%
      \csname LTb\endcsname%
      \put(946,1733){\makebox(0,0)[r]{\strut{}$9$}}%
      \csname LTb\endcsname%
      \put(946,2247){\makebox(0,0)[r]{\strut{}$9.5$}}%
      \csname LTb\endcsname%
      \put(946,2762){\makebox(0,0)[r]{\strut{}$10$}}%
      \csname LTb\endcsname%
      \put(946,3276){\makebox(0,0)[r]{\strut{}$10.5$}}%
      \csname LTb\endcsname%
      \put(946,3790){\makebox(0,0)[r]{\strut{}$11$}}%
      \csname LTb\endcsname%
      \put(946,4305){\makebox(0,0)[r]{\strut{}$11.5$}}%
      \csname LTb\endcsname%
      \put(946,4819){\makebox(0,0)[r]{\strut{}$12$}}%
      \csname LTb\endcsname%
      \put(1078,484){\makebox(0,0){\strut{}$0$}}%
      \csname LTb\endcsname%
      \put(2509,484){\makebox(0,0){\strut{}$50$}}%
      \csname LTb\endcsname%
      \put(3941,484){\makebox(0,0){\strut{}$100$}}%
      \csname LTb\endcsname%
      \put(5372,484){\makebox(0,0){\strut{}$150$}}%
      \csname LTb\endcsname%
      \put(6803,484){\makebox(0,0){\strut{}$200$}}%
    }%
    \gplgaddtomacro\gplfronttext{%
      \csname LTb\endcsname%
      \put(5816,4591){\makebox(0,0)[r]{\strut{}$\sigma=0.1\sigma_0$}}%
      \csname LTb\endcsname%
      \put(5816,4261){\makebox(0,0)[r]{\strut{}$\sigma=\sigma_0$}}%
      \csname LTb\endcsname%
      \put(5816,3931){\makebox(0,0)[r]{\strut{}$\sigma=10\sigma_0$}}%
      \csname LTb\endcsname%
      \put(209,2761){\rotatebox{-270.00}{\makebox(0,0){\strut{}$r$}}}%
      \put(3940,154){\makebox(0,0){\strut{}$t$}}%
    }%
    \gplbacktext
    \put(0,0){\includegraphics[width={360.00bp},height={252.00bp}]{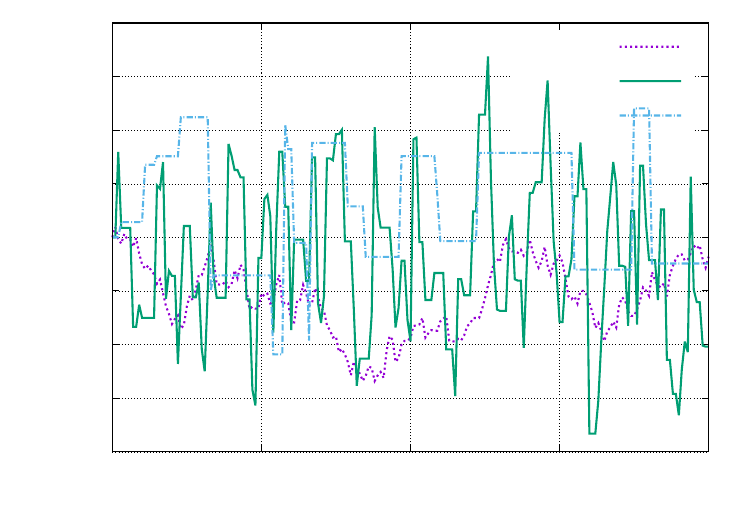}}%
    \gplfronttext
  \end{picture}%
\endgroup
\begingroup
  \inputencoding{latin1}%
  \makeatletter
  \providecommand\color[2][]{%
    \GenericError{(gnuplot) \space\space\space\@spaces}{%
      Package color not loaded in conjunction with
      terminal option `colourtext'%
    }{See the gnuplot documentation for explanation.%
    }{Either use 'blacktext' in gnuplot or load the package
      color.sty in LaTeX.}%
    \renewcommand\color[2][]{}%
  }%
  \providecommand\includegraphics[2][]{%
    \GenericError{(gnuplot) \space\space\space\@spaces}{%
      Package graphicx or graphics not loaded%
    }{See the gnuplot documentation for explanation.%
    }{The gnuplot epslatex terminal needs graphicx.sty or graphics.sty.}%
    \renewcommand\includegraphics[2][]{}%
  }%
  \providecommand\rotatebox[2]{#2}%
  \@ifundefined{ifGPcolor}{%
    \newif\ifGPcolor
    \GPcolortrue
  }{}%
  \@ifundefined{ifGPblacktext}{%
    \newif\ifGPblacktext
    \GPblacktexttrue
  }{}%
  \let\gplgaddtomacro\g@addto@macro
  \gdef\gplbacktext{}%
  \gdef\gplfronttext{}%
  \makeatother
  \ifGPblacktext
    \def\colorrgb#1{}%
    \def\colorgray#1{}%
  \else
    \ifGPcolor
      \def\colorrgb#1{\color[rgb]{#1}}%
      \def\colorgray#1{\color[gray]{#1}}%
      \expandafter\def\csname LTw\endcsname{\color{white}}%
      \expandafter\def\csname LTb\endcsname{\color{black}}%
      \expandafter\def\csname LTa\endcsname{\color{black}}%
      \expandafter\def\csname LT0\endcsname{\color[rgb]{1,0,0}}%
      \expandafter\def\csname LT1\endcsname{\color[rgb]{0,1,0}}%
      \expandafter\def\csname LT2\endcsname{\color[rgb]{0,0,1}}%
      \expandafter\def\csname LT3\endcsname{\color[rgb]{1,0,1}}%
      \expandafter\def\csname LT4\endcsname{\color[rgb]{0,1,1}}%
      \expandafter\def\csname LT5\endcsname{\color[rgb]{1,1,0}}%
      \expandafter\def\csname LT6\endcsname{\color[rgb]{0,0,0}}%
      \expandafter\def\csname LT7\endcsname{\color[rgb]{1,0.3,0}}%
      \expandafter\def\csname LT8\endcsname{\color[rgb]{0.5,0.5,0.5}}%
    \else
      \def\colorrgb#1{\color{black}}%
      \def\colorgray#1{\color[gray]{#1}}%
      \expandafter\def\csname LTw\endcsname{\color{white}}%
      \expandafter\def\csname LTb\endcsname{\color{black}}%
      \expandafter\def\csname LTa\endcsname{\color{black}}%
      \expandafter\def\csname LT0\endcsname{\color{black}}%
      \expandafter\def\csname LT1\endcsname{\color{black}}%
      \expandafter\def\csname LT2\endcsname{\color{black}}%
      \expandafter\def\csname LT3\endcsname{\color{black}}%
      \expandafter\def\csname LT4\endcsname{\color{black}}%
      \expandafter\def\csname LT5\endcsname{\color{black}}%
      \expandafter\def\csname LT6\endcsname{\color{black}}%
      \expandafter\def\csname LT7\endcsname{\color{black}}%
      \expandafter\def\csname LT8\endcsname{\color{black}}%
    \fi
  \fi
    \setlength{\unitlength}{0.0500bp}%
    \ifx\gptboxheight\undefined%
      \newlength{\gptboxheight}%
      \newlength{\gptboxwidth}%
      \newsavebox{\gptboxtext}%
    \fi%
    \setlength{\fboxrule}{0.5pt}%
    \setlength{\fboxsep}{1pt}%
    \definecolor{tbcol}{rgb}{1,1,1}%
\begin{picture}(7200.00,5040.00)%
    \gplgaddtomacro\gplbacktext{%
      \csname LTb\endcsname%
      \put(946,704){\makebox(0,0)[r]{\strut{}$-0.2$}}%
      \csname LTb\endcsname%
      \put(946,1390){\makebox(0,0)[r]{\strut{}$0$}}%
      \csname LTb\endcsname%
      \put(946,2076){\makebox(0,0)[r]{\strut{}$0.2$}}%
      \csname LTb\endcsname%
      \put(946,2762){\makebox(0,0)[r]{\strut{}$0.4$}}%
      \csname LTb\endcsname%
      \put(946,3447){\makebox(0,0)[r]{\strut{}$0.6$}}%
      \csname LTb\endcsname%
      \put(946,4133){\makebox(0,0)[r]{\strut{}$0.8$}}%
      \csname LTb\endcsname%
      \put(946,4819){\makebox(0,0)[r]{\strut{}$1$}}%
      \csname LTb\endcsname%
      \put(1078,484){\makebox(0,0){\strut{}$0$}}%
      \csname LTb\endcsname%
      \put(2223,484){\makebox(0,0){\strut{}$5$}}%
      \csname LTb\endcsname%
      \put(3368,484){\makebox(0,0){\strut{}$10$}}%
      \csname LTb\endcsname%
      \put(4513,484){\makebox(0,0){\strut{}$15$}}%
      \csname LTb\endcsname%
      \put(5658,484){\makebox(0,0){\strut{}$20$}}%
      \csname LTb\endcsname%
      \put(6803,484){\makebox(0,0){\strut{}$25$}}%
    }%
    \gplgaddtomacro\gplfronttext{%
      \csname LTb\endcsname%
      \put(5816,4591){\makebox(0,0)[r]{\strut{}$\sigma=0.5\sigma_0$}}%
      \csname LTb\endcsname%
      \put(5816,4261){\makebox(0,0)[r]{\strut{}$\sigma=\sigma_0$}}%
      \csname LTb\endcsname%
      \put(5816,3931){\makebox(0,0)[r]{\strut{}$\sigma=2\sigma_0$}}%
      \csname LTb\endcsname%
      \put(5816,3601){\makebox(0,0)[r]{\strut{}$\sigma=5\sigma_0$}}%
      \csname LTb\endcsname%
      \put(209,2761){\rotatebox{-270.00}{\makebox(0,0){\strut{}$C_{r}(\Delta t)$}}}%
      \put(3940,154){\makebox(0,0){\strut{}$\Delta t$}}%
    }%
    \gplbacktext
    \put(0,0){\includegraphics[width={360.00bp},height={252.00bp}]{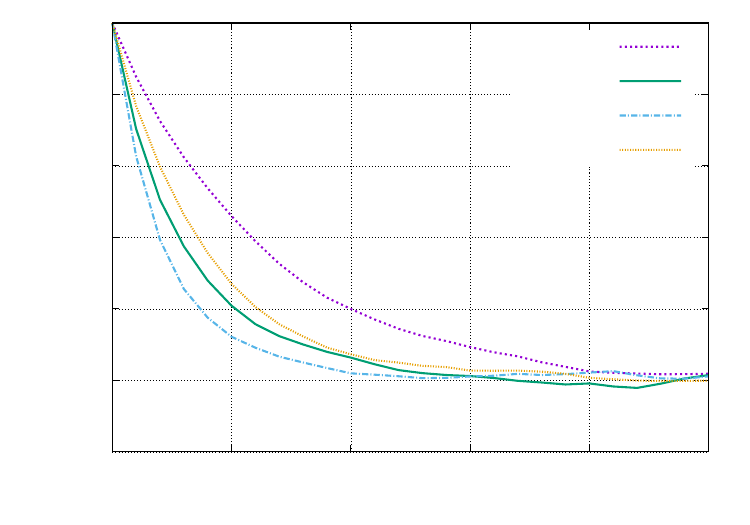}}%
    \gplfronttext
  \end{picture}%
\endgroup
}}
		\caption{Time series (left) and autocorrelation functions (right) of Markov chains generated using \cref{alg:poly_pot} based on \cref{th:dbl_exp_subst,th:updates_in_practice} with different standard deviations $\sigma$ in the update of $z\rightarrow z+\gamma$ with $\gamma\sim\mathcal{N}(0,\sigma^2)$. The potential $V(x)=\frac12 |x|^2$ (i.e.\ $p(r)\propto r^{d-1}\eto{-\frac12r^2}$) of the radius $r=\eto{z}$ in $d=100$ dimensions (i.e.\ $V_\text{eff}(z)=\frac12\eto{2z}-dz$) was treated as a one-dimensional problem and the angular component was not sampled for simplicity. Prediction~\eqref{eq:best_sigma} was used for $\sigma_0=\frac{1}{10}$.}\label{fig:time_series_autocorr}
	\end{figure}
	
	In order to verify the prediction~\eqref{eq:best_sigma} from \cref{th:scaling_sigma}, the probability distribution generated by the potential $V(r)=\frac12 r^2$ has been sampled in different Euclidean dimensions $d$. For simplicity, the angular component was neglected, that is we sampled $r\in[0,\infty)$ directly from the effective radial distribution $p(r)\propto r^{d-1}\eto{-\frac12r^2}$. As prescribed by \cref{th:updates_in_practice}, we used the substitution $r=\eto{z}$ from equation~\eqref{eq:subst_poly_pot}. The update $z\mapsto z + \gamma$ with $\gamma\sim \mathcal{N}(0,\sigma^2)$ then guarantees exponentially converging accurate sampling (see also \cref{th:gauss_update}) for any $\sigma>0$.
	
	Several time series for $d=100$ have been visualised in figure~\ref{fig:time_series_autocorr} (left) together with related autocorrelation functions $C_r(\Delta t)$ (right) for different choices of $\sigma$. From prediction~\eqref{eq:best_sigma} we expect some $\sigma\approx\sigma_0=\frac{1}{10}$ to minimise the autocorrelation. As expected, a much smaller standard deviation of $\sigma=\num{0.1}\sigma_0$ leads to small steps in the time series while a much larger choice of $\sigma=10\sigma_0$ results in very low acceptance rates. Both extremes clearly lead to long autocorrelation times. The right hand side panel of figure~\ref{fig:time_series_autocorr} indicates that $\sigma\approx2\sigma_0$ leads to even better decorrelation, though $\sigma=\sigma_0$ already produces close to optimal results. It is not surprising that we find a deviation of order 1 from the predicted value since the calculation of $\sigma_0$ relies on \cref{th:max_decorr} which is not exact.
	
	\begin{figure}[th]
		\centering
		\resizebox{0.98\textwidth}{!}{{\large%
\begingroup
  \inputencoding{latin1}%
  \makeatletter
  \providecommand\color[2][]{%
    \GenericError{(gnuplot) \space\space\space\@spaces}{%
      Package color not loaded in conjunction with
      terminal option `colourtext'%
    }{See the gnuplot documentation for explanation.%
    }{Either use 'blacktext' in gnuplot or load the package
      color.sty in LaTeX.}%
    \renewcommand\color[2][]{}%
  }%
  \providecommand\includegraphics[2][]{%
    \GenericError{(gnuplot) \space\space\space\@spaces}{%
      Package graphicx or graphics not loaded%
    }{See the gnuplot documentation for explanation.%
    }{The gnuplot epslatex terminal needs graphicx.sty or graphics.sty.}%
    \renewcommand\includegraphics[2][]{}%
  }%
  \providecommand\rotatebox[2]{#2}%
  \@ifundefined{ifGPcolor}{%
    \newif\ifGPcolor
    \GPcolortrue
  }{}%
  \@ifundefined{ifGPblacktext}{%
    \newif\ifGPblacktext
    \GPblacktexttrue
  }{}%
  \let\gplgaddtomacro\g@addto@macro
  \gdef\gplbacktext{}%
  \gdef\gplfronttext{}%
  \makeatother
  \ifGPblacktext
    \def\colorrgb#1{}%
    \def\colorgray#1{}%
  \else
    \ifGPcolor
      \def\colorrgb#1{\color[rgb]{#1}}%
      \def\colorgray#1{\color[gray]{#1}}%
      \expandafter\def\csname LTw\endcsname{\color{white}}%
      \expandafter\def\csname LTb\endcsname{\color{black}}%
      \expandafter\def\csname LTa\endcsname{\color{black}}%
      \expandafter\def\csname LT0\endcsname{\color[rgb]{1,0,0}}%
      \expandafter\def\csname LT1\endcsname{\color[rgb]{0,1,0}}%
      \expandafter\def\csname LT2\endcsname{\color[rgb]{0,0,1}}%
      \expandafter\def\csname LT3\endcsname{\color[rgb]{1,0,1}}%
      \expandafter\def\csname LT4\endcsname{\color[rgb]{0,1,1}}%
      \expandafter\def\csname LT5\endcsname{\color[rgb]{1,1,0}}%
      \expandafter\def\csname LT6\endcsname{\color[rgb]{0,0,0}}%
      \expandafter\def\csname LT7\endcsname{\color[rgb]{1,0.3,0}}%
      \expandafter\def\csname LT8\endcsname{\color[rgb]{0.5,0.5,0.5}}%
    \else
      \def\colorrgb#1{\color{black}}%
      \def\colorgray#1{\color[gray]{#1}}%
      \expandafter\def\csname LTw\endcsname{\color{white}}%
      \expandafter\def\csname LTb\endcsname{\color{black}}%
      \expandafter\def\csname LTa\endcsname{\color{black}}%
      \expandafter\def\csname LT0\endcsname{\color{black}}%
      \expandafter\def\csname LT1\endcsname{\color{black}}%
      \expandafter\def\csname LT2\endcsname{\color{black}}%
      \expandafter\def\csname LT3\endcsname{\color{black}}%
      \expandafter\def\csname LT4\endcsname{\color{black}}%
      \expandafter\def\csname LT5\endcsname{\color{black}}%
      \expandafter\def\csname LT6\endcsname{\color{black}}%
      \expandafter\def\csname LT7\endcsname{\color{black}}%
      \expandafter\def\csname LT8\endcsname{\color{black}}%
    \fi
  \fi
    \setlength{\unitlength}{0.0500bp}%
    \ifx\gptboxheight\undefined%
      \newlength{\gptboxheight}%
      \newlength{\gptboxwidth}%
      \newsavebox{\gptboxtext}%
    \fi%
    \setlength{\fboxrule}{0.5pt}%
    \setlength{\fboxsep}{1pt}%
    \definecolor{tbcol}{rgb}{1,1,1}%
\begin{picture}(7200.00,5040.00)%
    \gplgaddtomacro\gplbacktext{%
      \csname LTb\endcsname%
      \put(946,1200){\makebox(0,0)[r]{\strut{}2.3}}%
      \csname LTb\endcsname%
      \put(946,704){\makebox(0,0)[r]{\strut{}$1$}}%
      \csname LTb\endcsname%
      \put(946,2076){\makebox(0,0)[r]{\strut{}$10$}}%
      \csname LTb\endcsname%
      \put(946,3447){\makebox(0,0)[r]{\strut{}$100$}}%
      \csname LTb\endcsname%
      \put(946,4819){\makebox(0,0)[r]{\strut{}$1000$}}%
      \csname LTb\endcsname%
      \put(2265,484){\makebox(0,0){\strut{}$0.01$}}%
      \csname LTb\endcsname%
      \put(4237,484){\makebox(0,0){\strut{}$0.1$}}%
      \csname LTb\endcsname%
      \put(6209,484){\makebox(0,0){\strut{}$1$}}%
    }%
    \gplgaddtomacro\gplfronttext{%
      \csname LTb\endcsname%
      \put(5816,4591){\makebox(0,0)[r]{\strut{}$d=10$, data}}%
      \csname LTb\endcsname%
      \put(5816,4261){\makebox(0,0)[r]{\strut{}$d=10$, fit}}%
      \csname LTb\endcsname%
      \put(5816,3931){\makebox(0,0)[r]{\strut{}$d=100$, data}}%
      \csname LTb\endcsname%
      \put(5816,3601){\makebox(0,0)[r]{\strut{}$d=100$, fit}}%
      \csname LTb\endcsname%
      \put(5816,3271){\makebox(0,0)[r]{\strut{}$d=1000$, data}}%
      \csname LTb\endcsname%
      \put(5816,2941){\makebox(0,0)[r]{\strut{}$d=1000$, fit}}%
      \csname LTb\endcsname%
      \put(209,2761){\rotatebox{-270.00}{\makebox(0,0){\strut{}$\tint$}}}%
      \put(3940,154){\makebox(0,0){\strut{}$\sigma$}}%
    }%
    \gplbacktext
    \put(0,0){\includegraphics[width={360.00bp},height={252.00bp}]{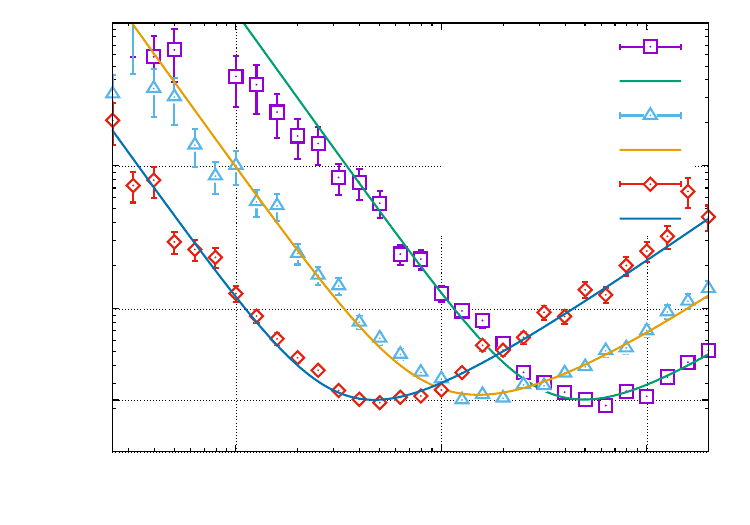}}%
    \gplfronttext
  \end{picture}%
\endgroup
\begingroup
  \inputencoding{latin1}%
  \makeatletter
  \providecommand\color[2][]{%
    \GenericError{(gnuplot) \space\space\space\@spaces}{%
      Package color not loaded in conjunction with
      terminal option `colourtext'%
    }{See the gnuplot documentation for explanation.%
    }{Either use 'blacktext' in gnuplot or load the package
      color.sty in LaTeX.}%
    \renewcommand\color[2][]{}%
  }%
  \providecommand\includegraphics[2][]{%
    \GenericError{(gnuplot) \space\space\space\@spaces}{%
      Package graphicx or graphics not loaded%
    }{See the gnuplot documentation for explanation.%
    }{The gnuplot epslatex terminal needs graphicx.sty or graphics.sty.}%
    \renewcommand\includegraphics[2][]{}%
  }%
  \providecommand\rotatebox[2]{#2}%
  \@ifundefined{ifGPcolor}{%
    \newif\ifGPcolor
    \GPcolortrue
  }{}%
  \@ifundefined{ifGPblacktext}{%
    \newif\ifGPblacktext
    \GPblacktexttrue
  }{}%
  \let\gplgaddtomacro\g@addto@macro
  \gdef\gplbacktext{}%
  \gdef\gplfronttext{}%
  \makeatother
  \ifGPblacktext
    \def\colorrgb#1{}%
    \def\colorgray#1{}%
  \else
    \ifGPcolor
      \def\colorrgb#1{\color[rgb]{#1}}%
      \def\colorgray#1{\color[gray]{#1}}%
      \expandafter\def\csname LTw\endcsname{\color{white}}%
      \expandafter\def\csname LTb\endcsname{\color{black}}%
      \expandafter\def\csname LTa\endcsname{\color{black}}%
      \expandafter\def\csname LT0\endcsname{\color[rgb]{1,0,0}}%
      \expandafter\def\csname LT1\endcsname{\color[rgb]{0,1,0}}%
      \expandafter\def\csname LT2\endcsname{\color[rgb]{0,0,1}}%
      \expandafter\def\csname LT3\endcsname{\color[rgb]{1,0,1}}%
      \expandafter\def\csname LT4\endcsname{\color[rgb]{0,1,1}}%
      \expandafter\def\csname LT5\endcsname{\color[rgb]{1,1,0}}%
      \expandafter\def\csname LT6\endcsname{\color[rgb]{0,0,0}}%
      \expandafter\def\csname LT7\endcsname{\color[rgb]{1,0.3,0}}%
      \expandafter\def\csname LT8\endcsname{\color[rgb]{0.5,0.5,0.5}}%
    \else
      \def\colorrgb#1{\color{black}}%
      \def\colorgray#1{\color[gray]{#1}}%
      \expandafter\def\csname LTw\endcsname{\color{white}}%
      \expandafter\def\csname LTb\endcsname{\color{black}}%
      \expandafter\def\csname LTa\endcsname{\color{black}}%
      \expandafter\def\csname LT0\endcsname{\color{black}}%
      \expandafter\def\csname LT1\endcsname{\color{black}}%
      \expandafter\def\csname LT2\endcsname{\color{black}}%
      \expandafter\def\csname LT3\endcsname{\color{black}}%
      \expandafter\def\csname LT4\endcsname{\color{black}}%
      \expandafter\def\csname LT5\endcsname{\color{black}}%
      \expandafter\def\csname LT6\endcsname{\color{black}}%
      \expandafter\def\csname LT7\endcsname{\color{black}}%
      \expandafter\def\csname LT8\endcsname{\color{black}}%
    \fi
  \fi
    \setlength{\unitlength}{0.0500bp}%
    \ifx\gptboxheight\undefined%
      \newlength{\gptboxheight}%
      \newlength{\gptboxwidth}%
      \newsavebox{\gptboxtext}%
    \fi%
    \setlength{\fboxrule}{0.5pt}%
    \setlength{\fboxsep}{1pt}%
\begin{picture}(7200.00,5040.00)%
    \gplgaddtomacro\gplbacktext{%
      \csname LTb\endcsname%
      \put(814,2465){\makebox(0,0)[r]{\strut{}$0.1$}}%
      \csname LTb\endcsname%
      \put(946,484){\makebox(0,0){\strut{}$10$}}%
      \csname LTb\endcsname%
      \put(3875,484){\makebox(0,0){\strut{}$100$}}%
      \csname LTb\endcsname%
      \put(6803,484){\makebox(0,0){\strut{}$1000$}}%
    }%
    \gplgaddtomacro\gplfronttext{%
      \csname LTb\endcsname%
      \put(209,2761){\rotatebox{-270}{\makebox(0,0){\strut{}$\sigma_\text{min}$}}}%
      \put(3874,154){\makebox(0,0){\strut{}$d$}}%
      \csname LTb\endcsname%
      \put(5816,4591){\makebox(0,0)[r]{\strut{}data}}%
      \csname LTb\endcsname%
      \put(5816,4261){\makebox(0,0)[r]{\strut{}prediction~\eqref{eq:best_sigma}}}%
      \csname LTb\endcsname%
      \put(5816,3931){\makebox(0,0)[r]{\strut{}fit}}%
    }%
    \gplbacktext
    \put(0,0){\includegraphics{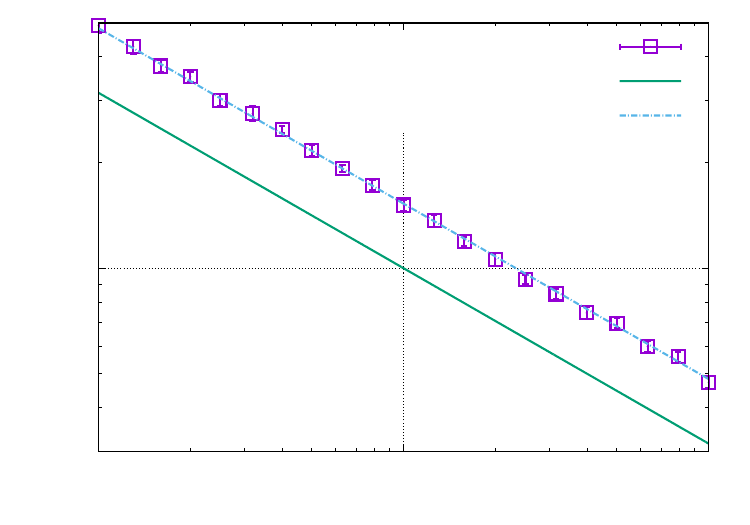}}%
    \gplfronttext
  \end{picture}%
\endgroup
}}
		\caption{Left: \intaut\ of Markov chains in the same setting as figure~\ref{fig:time_series_autocorr} as a function of the standard deviation $\sigma$ for different dimensions $d$. The fit function has the form $\tint = a\sigma^{-2} + b\sigma + c$. Right: Minima $\sigma_\text{min}=\left(\frac ab\right)^{1/3}$ obtained from the fits of $\tint$. Prediction~\eqref{eq:best_sigma} suggests $\sigma_\text{min}=\frac{1}{\sqrt d}$ while a fit $\sigma_\text{min}=\frac{\sigma^*}{\sqrt d}$ yields $\sigma^*=\num{1.528\pm 0.007}$.}\label{fig:opt_sigma}
	\end{figure}
	
	These deviations are quantified more precisely in figure~\ref{fig:opt_sigma}. Therein the \intaut\ $\tint$ is plotted for different dimensions $d$ as a function of the standard deviation $\sigma$ in the left panel. For very small $\sigma$ the update is expected to enter a diffusive regime where the autocorrelation scales as $\tint\propto\sigma^{-2}$ while for very large $\sigma$ the acceptance decreases linearly leading to $\tint\propto\sigma$. These asymptotic considerations together with a constant off-set motivate the fitting ansatz $\tint = a\sigma^{-2} + b\sigma + c$. The data is very well described by this ansatz in a wide region about the minimum of $\tint$ as can be seen in the left panel of figure~\ref{fig:opt_sigma}. The fit allows to extract the optimal standard deviation $\sigma_\text{min}\left(\frac ab\right)^{1/3}$ for each dimension. The respective results are visualised in the right panel of figure~\ref{fig:opt_sigma}. Again, we find that the prediction~\eqref{eq:best_sigma} is not exact as $\sigma_\text{min}$ deviates from the expected value $1/\sqrt{d}$ by a constant factor $\sigma*=\num{1.528\pm 0.007}$. Nevertheless, the main result of \cref{th:scaling_sigma}, namely the scaling with the inverse square root of the dimension, could be confirmed.
	
	Finally, let us check the predictions in \cref{th:opt_acc} since they are also based on the \cref{th:max_decorr}. \Cref{th:opt_acc} predicts the minimal \intaut\ $\tint \approx \num{2.3}$ for the radial variable $r$ and an acceptance rate $p_\text{acc} \gtrsim\num{0.42}$. From the left panel of figure~\ref{fig:opt_sigma} it is apparent that the estimator of the minimal $\tint$ is even quantitatively surprisingly accurate. Note that once the \radup\ is combined with another update, this value might decrease. But it should never be much larger unless the potential is much more intricate. Note also that other observables can have significantly deviating autocorrelations from that of $r$.
	
	The easiest way to tune $\sigma$ for more complicated potentials in practice is by targeting a given acceptance rate. $\sigma=\sigma_0$ can be used as an initial guess, but it can be off by some non-negligible factor as we have seen. It turns out that in the simulations at hand the minimal autocorrelation was achieved at an acceptance rate of $p_\text{acc} =\num{0.482\pm0.005}$, practically independently of the dimension $d$. This is in good agreement with the expectation $p_\text{acc} \gtrsim\num{0.42}$. Therefore, a target acceptance rate of (easily memorable) $50\%$ is a good choice.
 	
	\section{Conclusion}\label{sec:conclusion}

In this work we have generalised the approach introduced in Ref.~\cite{original_radial_update} to use \radup s for the efficient sampling of probability distributions defined on non-compact spaces. The idea as explained in \cref{sec:algorithm} is to split the configuration into a radial and an angular part (we call spaces where this is possible \radman s). Then the \radup\ samples the radial component with appropriately scaled step sizes that allow quick movement through the phase space. The \radup\ has to be combined with another Markov chain Monte Carlo (MCMC) method that takes care of the angular component.

The realisation in practice has been summarised in \cref{alg:poly_pot} for polynomial potentials and in \cref{alg:any_pot} in the general case. We have also demonstrated in \cref{sec:examples} that these algorithms allow reliable sampling, even from very heavy-tailed distributions that are intractable with conventional MCMC sampling.
Moreover, \radup s combined with the Hybrid Monte Carlo (HMC) algorithm have proven greatly successful in simulations of realistic physical systems beyond toy models~\cite{Temmen:2024pcm,Temmen_ergodicity}. 

We formalise the concept of the combined algorithm in \cref{th:hmc_convergence_non-compact}. It states that, given an MCMC algorithm that samples every compact environment correctly, the combination of this algorithm with an appropriately chosen \radup\ guarantees exponential convergence to the target probability distribution $p(x)$. \Cref{th:dbl_exp_subst} then specifies this appropriate choice: the radial variable $r=||x||$ is substituted $r=f(z)$ so that the \effpot\ $V_\text{eff}$ defining $p(z,\theta)\propto \eto{-V_\text{eff}(z,\theta)}$ grows exponentially in the auxiliary variable $z$. Then the \radup\ is performed by local additive steps in said variable $z$. Some substitutions that can be useful in practice are listed in \cref{th:updates_in_practice}.
	
A further generalisation is provided in \cref{th:classification}. Therein we find that even badly chosen \radup s will lead to convergence and their convergence speed is classified. It follows in \cref{th:hmc_conv} that the hybrid Monte Carlo (HMC) algorithm is guaranteed to converge even without radial update. However, it cannot be emphasised enough that a well-chosen radial update can accelerate the convergence by many orders of magnitude and the other variations listed in \cref{th:classification} should therefore never be used in practice.

Finally, for \cref{th:scaling_sigma} we derive the optimal parameter choice for the \radup\ in the case of asymptotically polynomial potentials. Given the potential $V(x)\approx c|x|^a$ for large $|x|$, in $d$ Euclidean dimensions the \radup\ should propose $x\mapsto x\eto{\gamma}$ with $\gamma\sim\mathcal{N}(0,\sigma^2)$ normally distributed and the standard deviation $\sigma\simeq \sqrt{\frac{2}{ad}}$. Here, the factor $\sqrt{2}$ might be slightly off from the optimal value, but the scaling in $a$ and $d$ is universal.

Overall, \radup s are a very powerful and computationally inexpensive tool. They should therefore be integrated in every MCMC simulation on non-compact \radman s. 	
	\section*{Code and Data}
	All simulations in this work were implemented in \texttt{R} and the code has been made publicly available~\cite{johann_ostmeyer_2024_14197590}.
	The analysis used the light-weight tool \texttt{comp-avg}~\cite{comp-avg}. The simulations can be reproduced very quickly, nonetheless the resulting data will gladly be provided upon request.
	
	\section*{Acknowledgements}
	The author thanks Anthony Kennedy and Xinhao Yu for highly insightful discussions, especially for pointing out flaws in the first versions of some proofs. Further thanks for valuable comments go to Evan Berkowitz, Marco Garofalo, Tom Luu, Finn Temmen, and Carsten Urbach.
	This work was funded in part by the Deutsche Forschungsgemeinschaft (DFG, German Research Foundation) as part of the CRC 1639 NuMeriQS -- project no.\ 511713970.

	\FloatBarrier
	\printbibliography

	\allowdisplaybreaks[1]
	\appendix

\section[Collection of proofs]{Collection of proofs including those of \cref{th:hmc_conv_compact,th:hmc_convergence_non-compact,th:really_hmc_convergence_non-compact,th:updates_in_practice,th:scaling_sigma,th:max_decorr,th:opt_acc}}\label{sec:proofs}

\begin{proof}[Proof from~\cite{original_radial_update}: Sufficiency of the \cdc, \cref{def:compact_doeblin}]
	 If it holds that
	 \begin{align}
	 \inf_{x\in \mathcal{C}} \mathcal{P}(x, y) &\ge \xi \nu(y)
	 \end{align}
	 for all $y\in\mathcal{C}$, a new probability measure $\tilde{\nu}$ can be defined such that
	 \begin{align}
	 	\tilde{\nu}(y) &= \begin{cases}
	 		\frac{1}{\nu(\mathcal{C})}\nu(y) & \text{if } y\in \mathcal{C}\\
	 		0 & \text{otherwise}
	 	\end{cases}
	 \end{align}
	 which is well defined since $\mathcal{C}$ is compact therefore $\nu(\mathcal{C})\in(0,1]$ is a bounded number, the traditional
	 Doeblin’s condition then holds with $\nu$ replaced by $\tilde{\nu}$ and $\xi$ replaced by $\xi\nu(\mathcal{C})>0$.
	 So we can always restrict ourselves to transitions starting and ending inside $\mathcal{C}$.
\end{proof}

\begin{proof}[Proof from~\cite{original_radial_update}: HMC satisfies \cdc, \cref{th:hmc_conv_compact}]
	Ref.~\cite{original_radial_update} contains a very detailed general proof of this statement, including the case of partial momentum refreshment. Here, instead we focus on the general idea of the proof.
	
	First, note that every new state proposed by the momentum refreshment and molecular dynamics step will have a non-zero acceptance probability. This follows from the Hamiltonian $H$ being continuous (the sum of the continuous potential and the smooth kinetic energy) on a compact set, thus reaching its finite maximum in modulus $-H_\text{max}\le H\le H_\text{max}$. Hence, the acceptance probability is bounded from below $p_\text{acc}\ge\exp\left(-\left|H-H'\right|\right)\ge\exp\left(-2H_\text{max}\right)>0$ for any two Hamiltonian terms $H$ and $H'$.
	
	It remains to show that every point within the compact set $y\in\mathcal{C}$ has a non-vanishing probability density $\inf_{x\in \mathcal{C}} \mathcal{P}(x, y) \ge \xi \nu(y)$ to be proposed as an update starting from any other point $x$ within $\mathcal{C}$. In fact, even the stronger statement is true that there exists some $\tilde\xi>0$ so that $\inf_{x,y\in \mathcal{C}} \mathcal{P}(x, y) \ge \tilde\xi$, i.e.\ a uniform probability density can be chosen.
	
	Intuitively, this is the case because the momenta can be arbitrarily large and there is always a set of momenta pointing in the required direction so as to reach the desired point. Since the potential is bounded, there are no unsurmountable potential barriers (no ergodicity violation). More rigorously, the Hopf–Rinow theorem is applicable to \radman s because they are connected and smooth as well as complete. This implies the existence of a minimal geodesic connecting any two points. Molecular dynamics follow a geodesic and therefore for any $x,y\in\mathcal{C}$ there exist momenta that allow the transition from $x$ to $y$ with non-vanishing probability density.
	
	Finally, partial momentum refreshment effectively rescales the magnitude of the random momenta by a non-zero factor. Therefore, the proof proceeds in exactly the same way but for an irrelevant rescaling of the effective trajectory length and consequently the probability bounds.
\end{proof}

\begin{proof}[Proof: Convergence with \radup, \cref{th:hmc_convergence_non-compact,th:really_hmc_convergence_non-compact}]
	 The proof of \cref{th:really_hmc_convergence_non-compact} originally presented in Ref.~\cite{original_radial_update} directly generalises to \cref{th:hmc_convergence_non-compact}. Here, we present an alternative more compact version, relying on the additional identities introduced in this work.
	 
	 First of all, we notice that the \sgdc\ is always satisfied on compact spaces according to \cref{th:conv_with_cdc_sgdc}. Therefore, $\mathcal{P}$ also satisfies the \sgdc\ on the compact $X_0$ and thus on the entire manifold $X$. From \cref{th:compositions} follows that the combined update with one component satisfying the \wgdc\ and the other the \sgdc\ satisfies the \sgdc. The combined algorithm also satisfies the \cdc\ since the \radup\ does not introduce any forbidden regions. Therefore, the combined algorithm satisfies all the requirements for exponential convergence as in \cref{th:conv_with_cdc_sgdc}.
	 
	 \Cref{th:really_hmc_convergence_non-compact} follows since \cref{th:hmc_conv_compact} guarantees that the HMC algorithm satisfies the \cdc\ while \cref{th:guarantee_of_wgdc} guarantees the \wgdc, thus fulfilling the requirements to apply \cref{th:hmc_convergence_non-compact}.
\end{proof}

\begin{proof}[Proof: (log-)normal \radup, \cref{th:updates_in_practice}]
	The updates listed above are asymptotically equivalent to substitutions satisfying condition~\eqref{eq:subst_condition}, so that \cref{th:dbl_exp_subst} is directly applicable. The explicit substitutions in $d$ dimensions are:
	\begin{itemize}
		\item exponential potential: $r = z\Rightarrow z+\gamma = r+\gamma,\ V_\text{eff}(z) = c\eto{a z+o(z)}-(d-1)\ln z$
		\item polynomial potential: $r = \eto{z}\Rightarrow \eto{z+\gamma} = r\eto{\gamma},\ V_\text{eff}(z) = c\eto{a z+o(z)} - dz$
		\item logarithmic potential: $r = \eto{\eto{z}}\Rightarrow \eto{\eto{z+\gamma}}=r^{\eto{\gamma}},\ V_\text{eff}(z) = (c-d)\eto{z+o(z)} - z$
	\end{itemize}
	Clearly the remaining terms in $\log z$ and $z$ are negligible in the $\eto{o(z)}$-sense.
	In the case of the logarithmic potential we have to assume $c>d$ for normalisability of the probability distribution. Under this assumption, condition~\eqref{eq:subst_condition} always holds.
	
	The acceptance probabilities $p_\text{acc}$ follow directly from the change of the respective effective potential $\Delta V_\text{eff}$.
\end{proof}

\begin{proof}[Proof: Scaling of log-normal $\sigma$ for polynomial potentials, \cref{th:scaling_sigma}]
	We match the log-normal distribution to the target distribution $p(r)\propto \eto{-V(r)}$. For this we first have to derive the mean $\mu_r$ and standard deviation $\sigma_r$ of $p(r)$. Then we apply \cref{th:max_decorr} to rescale $\sigma_r$ to the optimal standard deviation of the proposal distribution $\sigma_\text{opt}=\sqrt{2}\sigma_r$. Finally, the parameter $\sigma$ can be calculated from the parameters the log-normal distribution is supposed to resemble
	\begin{align*}
		\sigma^2 &= \ln\left(1+\frac{\sigma_\text{opt}^2}{\mu_r^2}\right)\\
		&=\ln\left(1+\frac{2}{ad}+\ordnung{d^{-2}}\right)\\
		&= \frac{2}{ad}+\ordnung{d^{-2}}\\
		\Rightarrow \sigma &= \sqrt{\frac{2}{ad}} + \ordnung{d^{-1}}\,.
	\end{align*}
	
	It remains to show that
	\begin{align}
		\frac{\sigma_r^2}{\mu_r^2} &= \frac{1}{ad}+\ordnung{d^{-2}}
	\end{align}
	indeed holds. Consider the integral
	\begin{align}
		I_s(c,a;d) &\coloneqq \int_{0}^{\infty}\md r r^{d-1+s}\eto{-cr^a}\,,
	\end{align}
	which encodes the normalisation of $p(r)$ for $s=0$, the mean for $s=1$ and the variance for $s=2$ in $d$-dimensional spherical coordinates. The asymptotic behaviour of $V(r)$ used in $I_s$ becomes an arbitrarily accurate approximation of the true expectation value with large enough $d$ because all contributions close to $r=0$ are exponentially suppressed in $d$. With the substitution $cr^a=z$ we can solve the integral exactly:
	\begin{align}
		I_s(c,a;d) &= \int_{0}^{\infty}\md z \frac{1}{ac^{\frac{d+s}{a}}} z^{\frac{d+s-a}{a}}\eto{-z}\\
		&= \frac{1}{ac^{\frac{d+s}{a}}} \Gamma\left(\frac{d+s}{a}\right)\,.
	\end{align}
	With the additional approximation~\cite{gamma_ratio_1951}
	\begin{align}
		\frac{\Gamma(x+y)}{\Gamma(x)} &= x^y + \frac{y(y-1)}{2}x^{y-1} + \ordnung{x^{y-2}}
	\end{align}
	we obtain
	\begin{alignat}{3}
		&\mu_r &&\equiv \frac{I_1(c,a;d)}{I_0(c,a;d)} &&= c^{-\frac1a} \left[\left(\frac da\right)^{\frac1a}+\frac{1}{2a}\left(\frac1a-1\right)\left(\frac da\right)^{\frac1a-1}\right] + \ordnung{d^{\frac 1a-2}}\,,\\
		&\sigma_r^2 &&\equiv \frac{I_2(c,a;d)}{I_0(c,a;d)}-\mu_r^2 &&= c^{-\frac2a} \left[\left(\frac da\right)^{\frac2a}+\frac1a\left(\frac2a-1\right)\left(\frac da\right)^{\frac2a-1}\right.\\
		& && && \qquad\left.-\left(\frac da\right)^{\frac2a}-\frac{1}{a}\left(\frac1a-1\right)\left(\frac da\right)^{\frac2a-1}\right] + \ordnung{d^{\frac 2a-2}}\\
		& && &&= c^{-\frac2a}\frac{1}{a^2}\left(\frac da\right)^{\frac2a-1} + \ordnung{d^{\frac 2a-2}}\\
		& &&\Rightarrow\frac{\sigma_r^2}{\mu_r^2} &&= \frac{1}{a^2}\left(\frac da\right)^{-1} + \ordnung{d^{-2}}\\
		& && &&=\frac{1}{ad}+\ordnung{d^{-2}}
	\end{alignat}
\end{proof}

\begin{proof}[Proof: Maximal decorrelation by additive update, \cref{th:max_decorr,th:opt_acc}]
	The proofs of \cref{th:max_decorr,th:opt_acc} proceed together. We first denote the probability density of $\gamma$ by $\rho(\gamma)$ and the acceptance probability of $z+\gamma$ by $A(z,z+\gamma)$ to estimate the covariance of $z$ and $z+\gamma$ (i.e.\ the decorrelation after a single step)
	\begin{align}
		\mathrm{cov}(z,z+\gamma) &= \int_{-\infty}^{\infty}\md\gamma \rho(\gamma) \left[(z-\mu)(z+\gamma-\mu)A(z,z+\gamma) + (z-\mu)^2(1-A(z,z+\gamma))\right]\\
		&= \int_{-\infty}^{\infty}\md\gamma \rho(\gamma) \left[(z-\mu)\gamma A(z,z+\gamma) + (z-\mu)^2\right]\\
		&=(z-\mu)^2+ (z-\mu)\int_{-\infty}^{\infty}\md\gamma \rho(\gamma) \gamma A(z,z+\gamma)\,.
	\end{align}
	Now we use that $p(z)$ decays quickly and therefore
	\begin{align}
		A(z,z+\gamma) &\simeq\begin{cases}
			1 & \text{if } |z+\gamma-\mu| \le |z-\mu|\,,\\
			0 & \text{else,}
		\end{cases}\label{eq:step_acc}
	\end{align}
	that is all updates towards the mean $\mu$ are accepted and all updates away from $\mu$ are rejected. The covariance integral then becomes solvable. W.l.o.g. let $z<\mu$, then
	\begin{align}
		\mathrm{cov}(z,z+\gamma) &\simeq (z-\mu)^2+ (z-\mu)\int_{0}^{2(\mu-z)}\md\gamma \rho(\gamma) \gamma\\
		&= (z-\mu)^2- (\mu-z)\frac{\sigma}{\sqrt{2\pi}}\left(1-\eto{-2\frac{(\mu-z)^2}{\sigma^2}}\right)\,.
	\end{align}
	For the expected correlation we can further approximately replace all occurrences of $\mu-z$ by $\sigma_z$ in order to obtain
	\begin{align}
		\erwartung{\mathrm{corr}(z,z+\gamma)} \equiv \frac{\erwartung{\mathrm{cov}(z,z+\gamma)}}{\sigma_z^2} &\simeq 1- \frac{\sigma}{\sigma_z\sqrt{2\pi}}\left(1-\eto{-2\frac{\sigma_z^2}{\sigma^2}}\right)\,.\label{eq:corr_z_z+gamma}
	\end{align}
	This correlation has a unique minimum in $\sigma$. The exact solution is a cumbersome poly-log expression and we have made quite radical approximations in \cref{eq:step_acc,eq:corr_z_z+gamma} already, which is why we restrain ourselves to the leading order (in large $\sigma$) solution
	\begin{align}
		\sigma &\simeq \sqrt{2}\sigma_z\,.
	\end{align}
	This completes the derivation of \cref{th:max_decorr}.
	
	Plugging the solution for $\sigma$ into the formula for the correlation~\eqref{eq:corr_z_z+gamma}, immediately yields equation~\eqref{eq:corr_z_gamma_opt}. The \intaut\ is then simply the geometric series
	\begin{align}
		\tint &= \frac12 + \sum_{k=1}^\infty \erwartung{\mathrm{corr}(z,z+\gamma)}^k\\
		&=\frac{1}{1-\erwartung{\mathrm{corr}(z,z+\gamma)}} - \frac12\\
		&\simeq \frac{\sigma_z\sqrt{2\pi}}{\sigma\left(1-\eto{-2\frac{\sigma_z^2}{\sigma^2}}\right)} - \frac12\\
		&= \frac{\sqrt{\pi}}{1-e^{-1}}-\frac12
	\end{align}
	and the acceptance rate in the scenario of rapid decay is
	\begin{align}
		p_\text{acc} &\gtrsim \int_{0}^{2(\mu-z)}\md\gamma \rho(\gamma)\\
		&= \frac12 \mathrm{erf}\left(\frac{2(\mu-z)}{\sqrt{2}\sigma}\right)\\
		&\simeq \frac12 \mathrm{erf}(1)\,.
	\end{align}
	
	Note that the correlation and therefore $\sigma$ as well as $\tint$ is not very sensitive to the approximation introduced in equation~\eqref{eq:step_acc} because it experiences both positive and negative corrections when $p(z)$ does not decay very rapidly. Detailed calculations with a more realistic model including an exponential tail of $A(z,z+\gamma)$ confirm this expectation. $p_\text{acc}$ on the other hand is bound from below by this approximation and can only be larger if the acceptance does not drop to 0 immediately for $|z+\gamma-\mu| > |z-\mu|$.
\end{proof}

\section{Proof of convergence for slower algorithms}\label{sec:slow_conv}

Here we introduce alternative drift conditions that are weaker than the \sgdc\ but still sufficient to guarantee convergence of an update scheme, be it slower than geometric. We subsequently generalise \cref{th:exp_pot}, and in consequence \cref{th:dbl_exp_subst}, and classify different convergence rates for various effective potentials.

\begin{definition}[Strong arithmetic drift condition]\label{def:sadc}
	An update scheme defined by the \mtk\ $\mathcal{P}$ satisfies the strong arithmetic drift condition (SADC) for the function $V:X\rightarrow[0,\infty)$ on the measurable space $X$ if some constant $K>0$ exist so that for all $x\in X$
	\begin{align}
		(\mathcal{P}V)(x) &\le V(x) - K\,.
	\end{align}
\end{definition}

\begin{remark}
	Analogously we can define the \textit{weak} arithmetic drift condition (WADC) if we allow $K\ge0$.
\end{remark}

\begin{definition}[Local arithmetic drift condition]\label{def:ladc}
An update scheme defined by the \mtk\ $\mathcal{P}$ satisfies the local arithmetic drift condition (LADC) for the function $V:X\rightarrow[0,\infty)$ on the measurable space $X$ if for all $x_0\in X$ some constant $K>0$ exist so that for all $x\in X$ with $V(x)\le V(x_0)$
\begin{align}
	(\mathcal{P}V)(x) &\le V(x) - K\,.
\end{align}
\end{definition}

\begin{proposition}[Hierarchy of drift conditions]\label{th:drift_hierarchy}
	Up to some compact region $X_0\subset X$ the drift conditions follow the hierarchy
	\begin{align}
		\text{\sgdc} \Rightarrow \text{\sadc} \Rightarrow \text{\ladc} \Rightarrow \text{\wadc} \Rightarrow \text{\wgdc.}
	\end{align}
\end{proposition}

\begin{proof}
	For $\text{\sgdc} \Rightarrow \text{\sadc}$ we use that $V(x)$ has to be larger than any given constant $K'$ everywhere but for some compact region $X_0$. Now choose $K'=\frac{K_g+K_a}{1-\alpha}$ and assume the \sgdc
	\begin{align}
		(\mathcal{P}V)(x) &\le \alpha V(x) + K_g\\
		&= V(x) - (1-\alpha) V(x) + K_g\\
		&\le V(x) - (1-\alpha) K' + K_g\\
		&= V(x) - K_a\,,
	\end{align}
	thus the \sadc\ follows.
	
	$\text{\sadc} \Rightarrow \text{\ladc}$ follows because the \sadc\ requires the existence of a global constant $K$ which automatically is a valid local constant everywhere as required by the \ladc.
	
	$\text{\ladc} \Rightarrow (\mathcal{P}V)(x) \le V(x) \Rightarrow \text{\wadc}$, simply setting $K=0$.
	
	$\text{\wadc} \Rightarrow (\mathcal{P}V)(x) \le V(x) \Rightarrow \text{\wgdc}$, simply setting $\alpha =0$ and $K=0$.
\end{proof}

\begin{lemma}[Guarantee of \wgdc]\label{th:guarantee_of_wgdc}
	An update scheme that uses a \symaker\ $\mathcal{P}$ (not necessarily additive) together with a \mhacc\ to sample from a probability distribution $p(x)\propto\eto{-V(x)}$ defined by the potential $V$ satisfies the \wgdc.
\end{lemma}

\begin{remark}
	Instead of the \mhacc\ one can also demand detailed balance. The proof then proceeds exactly analogously.
\end{remark}

\begin{proof}
	We start with the generalised version of equation~\eqref{eq:def_pav_1d} for the \radman\ $X$ rather then the 1D case	
	\begin{align}
		(\mathcal{P}_AV)(x) &= \int_X\md x'\mathcal{P}(x,x')\left[A(x,x')V(x') + \left(1-A(x,x')\right)V(x)\right]\\
		&= V(x) + \int_X\md x'\mathcal{P}(x,x')A(x,x')\left[V(x')-V(x)\right]\,.
	\end{align}
	Since the \mtk\ $\mathcal{P}$ is symmetric, the Metropolis-Hastings acceptance probability reduces to
	\begin{align}
		A(x,x') &= \min\left[1, \exp\left(-\left(V(x')-V(x)\right)\right)\right]\\
		 &= \begin{cases}
			1 & \text{if } x'\in X_0\,,\\
			\exp\left(-\left(V(x')-V(x)\right)\right) & \text{else,}
		\end{cases}
	\end{align}
	where $X_0\coloneqq \left\{x'\in X: V(x')\le V(x)\right\}$. Thus, the integral can be split into a negative part and a small part	
	\begin{align}
		(\mathcal{P}_AV)(x) &= V(x) + \int_{X_0}\!\!\md x'\mathcal{P}(x,x')\left[V(x')-V(x)\right] + \int_{X\setminus X_0}\!\!\!\!\md x'\mathcal{P}(x,x')\eto{-\left(V(x')-V(x)\right)}\left[V(x')-V(x)\right]\\
		&\le V(x) + \int_{X\setminus X_0}\!\!\!\!\md x'\mathcal{P}(x,x')\frac 1e\\
		&\le V(x) + \frac1e\,.
	\end{align}
\end{proof}

\begin{conjecture}[Guarantee of \ladc]\label{th:guarantee_of_ladc}
	Any update scheme on a \radman\ $X$ with the stationary probability distribution $p(x)\propto\eto{-V(x)}$ that updates the radial component in a \nontriv\ manner satisfies the \ladc\ for the potential $V$ on $X\setminus X_0$ with some compact $X_0$.
\end{conjecture}

\begin{remark}
	This conjecture is stronger than \cref{th:guarantee_of_wgdc} both in its assumptions and in its implication. If it holds, then \cref{th:guarantee_of_wgdc} immediately follows as a corollary. Now, the \cdc\ guarantees sampling everywhere in a compact region surrounding the current position, thus the radial direction must be explored in a \nontriv\ manner. Therefore, together with \cref{th:rates_of_conv} this conjecture would imply that every update scheme satisfying the \cdc\ automatically converges.
\end{remark}

\begin{argument}
	So far, we have no rigorous proof of this statement. Therefore, we only provide a motivation why we believe it to hold. Suppose the \wadc\ was not to hold even asymptotically. Then the expectation value of $V(x)$ would increase with every application of the update and eventually it would diverge, contradicting the assumption of a stationary distribution. The \nontriv\ update of the radial component motivates the stronger statement of \ladc\ as in \cref{th:diffus_conv} rather than the \wadc.
\end{argument}

\begin{lemma}[Composition of algorithms]\label{th:compositions}
	A composite algorithm satisfies the
	$\begin{Bmatrix}
		\text{\sgdc} \\ \text{\sadc / \ladc}
	\end{Bmatrix}$
	if each of its components satisfies the
	$\begin{Bmatrix}
		\text{\wgdc} \\ \text{\wadc}
	\end{Bmatrix}$
	and at least one component satisfies the
	$\begin{Bmatrix}
		\text{\sgdc} \\ \text{\sadc / \ladc}
	\end{Bmatrix}$.
\end{lemma}

\begin{remark}
	\Cref{th:compositions} generalises the reason why the HMC in combination with an update scheme that satisfies the \sgdc\ converges geometrically to the correct probability distribution. More generally, \cref{th:compositions} is one of the prerequisites for \cref{th:hmc_convergence_non-compact}.
\end{remark}

\begin{proof}
	The \wadc\ implies $(\mathcal{P}V)(x) \le V(x)$ and can thus be interpreted as a neutral element of algorithm composition. The identities for the \sadc\ and the \ladc\ follow immediately.
	
	The identity for the \sgdc\ is well known and has been used for instance in the proof of \cref{th:hmc_convergence_non-compact} in~\cite{original_radial_update}. It follows directly from the maximal increase of the Lyapunov function under the \wgdc\ by an additive constant. So the composite algorithm satisfies the \sgdc\ with a new constant $K=\sum_iK_i$ that is the sum of all component constants $K_i$.
\end{proof}

\begin{lemma}[Convergence equivalence]\label{th:same_convergence}
	If the Lyapunov function shrinks at a given rate (dictated by one of the drift conditions), then the distance between the probability distribution the target distribution shrinks at the same rate.
\end{lemma}

\begin{remark}
	In the case of geometric convergence \cref{th:same_convergence} has been derived in~\cite{hairer2008look} and it is another prerequisite for the convergence in \cref{th:hmc_convergence_non-compact}.
\end{remark}

\begin{proof}
	Most steps required for this proof are detailed in Ref.~\cite{hairer2008look} and we are not going to repeat them here. The only crucial difference for \cref{th:same_convergence} is the point where geometric convergence is explicitly assumed in~\cite{hairer2008look}. The equation proved in~\cite{hairer2008look} as a consequence of the \sgdc\ is
	\begin{align}
		|\mathcal{P}\varphi(x)-\mathcal{P}\varphi(y)| &\le \alpha d_\beta(x,y)
	\end{align}
	with $\alpha\in(0,1)$, $\beta>0$, some initial distribution $\varphi$ and the metric
	\begin{align}
		d_\beta(x,y)&=\begin{cases}
			0 & \text{if } x=y\,,\\
			2+\beta V(x) + \beta V(y) & \text{else.}
		\end{cases}
	\end{align}
	
	Now, assuming some more general form of reduction (any type of drift condition)
	\begin{align}
		(\mathcal{P}V)(x) &\le V(x) - \Delta(V(x), x)
	\end{align}
	with $\Delta: [0,\infty)\times X\rightarrow [0,\infty)$, we write
	\begin{align}
		|\mathcal{P}\varphi(x)-\mathcal{P}\varphi(y)| &\le 2 + \beta (\mathcal{P}V)(x) + \beta (\mathcal{P}V)(y)\\
		&\le 2 + \beta V(x) + \beta V(y) - \beta\left(\Delta(V(x), x)+ \Delta(V(y), y)\right)\\
		&= d_\beta(x,y) - \beta\left(\Delta(V(x), x)+ \Delta(V(y), y)\right)
	\end{align}
	and thus the distance of the distributions is reduced in the same rate. (In particular, for $\Delta(V(x),x)\propto V(x)$ we regain geometric reduction and for $\Delta(V(x),x)=\text{const.}$ we obtain arithmetic reduction.)
\end{proof}

\begin{definition}[Asymptotic approach]\label{def:approach}
	The approach of an update scheme defined by the \mtk\ $\mathcal{P}$ on the \radman\ $X$ is called
	$\begin{Bmatrix}
		\text{exponential} \\ \text{polynomial} \\ \text{diffusive}
	\end{Bmatrix}$
	if for $x\in X$ with asymptotically large $r\equiv ||x||$ the potential $V$ decreases at this rate, that is $(\mathcal{P}V)(x)< V(x)$ and
	$\begin{Bmatrix}
		\ln V(x) - \ln (\mathcal{P}V)(x) \in \Omega(1) \\ V(x) - (\mathcal{P}V)(x) \in \Omega(1) \text{ and } \ln V(x) - \ln (\mathcal{P}V)(x) \in o(1) \\ V(x) - (\mathcal{P}V)(x) \in o(1)
	\end{Bmatrix}$.
\end{definition}

\begin{lemma}[Convergence rates]\label{th:rates_of_conv}
	Let an update scheme on a \radman\ $X$ that satisfy the \cdc\ everywhere and the
	$\begin{Bmatrix}
		\text{\sgdc} \\ \text{\sadc} \\ \text{\ladc}
	\end{Bmatrix}$
	on $X\setminus X_0$ with some compact $X_0$. Then it converges exponentially to the target probability distribution $p(x)\propto\eto{-V(x)}$ given a continuous potential $V$ and it \approach es the region of exponential convergence at least
	$\begin{Bmatrix}
		\text{exponentially} \\ \text{polynomially} \\ \text{diffusively}
	\end{Bmatrix}$.
\end{lemma}

\begin{remark}
	The \wadc\ (and therefore the \wgdc) is not sufficient for convergence because it allows for the case $(\mathcal{P}V)(x) = V(x)$.
\end{remark}

\begin{proof}
	On a compact set \cref{th:hmc_convergence_non-compact} immediately guarantees exponential convergence. We are therefore going to show that the given drift condition implies that there exists a compact subset $X'\subset X$ which is \approach ed at the corresponding rate and never left in expectation value, i.e.\ if $x\in X'$ then $\mathcal{P}x\in X'$. Exponential convergence within $X'$ follows and \cref{th:same_convergence} guarantees that the probability distribution \approach es the target at the same rate.
	
	Let $R$ be a radius so that $V(x)\le R$ for all $x\in X_0$. Such an $R$ exists because $X_0$ is compact and $V$ continuous, thus $V$ reaches a finite maximum on $X_0$. For the same reason the Lyapunov function after a single update $(\mathcal{P}V)(x)\le R'$ is bounded when starting from $x\in X_0$. We set the compact `small set' $X'=\left\{x\in X: V(x) \le \max(R,R')\right\}\supset X_0$.
	
	Then $X'$ is \approach ed and never left as required for the exponential convergence. The \approach\ to $X'$ can be thought of as follows: for any starting point $x\in X\setminus X_0$ (including $x\in X'\setminus X_0$) the drift condition forces $\mathcal{P}^nx\in X_0\subset X'$ after some finite $n$. In particular, $(\mathcal{P}V)(x)\le V(x)$ and therefore $x\in X'$ implies $\mathcal{P}x\in X'$. For any starting point $x\in X_0$ the radius $R'$ was chosen such that $\mathcal{P}x\in X'$.
	
	It thus remains to show that the respective drift condition indeed guarantees that $X_0$ is \approach ed. 
	
	The \sgdc\ case is known from~\cite{hairer2008look} and the \sadc\ case follows analogously. More specifically, since $V$ is continuous, it is finite everywhere and a reduction by the fixed step $K$ dictated by the \sadc\ will reduce it below $R$ after a finite number $n$ of iterations.
	
	Once a finite starting point $x_0$ is fixed, the \ladc\ implies that the \sadc\ holds for all $x$ with $V(x)\le V(x_0)$ and the Lyapunov function $V$ cannot increase. Thus, the \ladc\ becomes equivalent to the \sadc, so convergence is guaranteed from every starting point. However, the convergence rate might depend on $x_0$ and, in principle, become arbitrarily slow and effectively indistinguishable from a random walk. Therefore, the convergence resembles diffusion rather than drift.
\end{proof}

\begin{lemma}[Optimality of geometric convergence]\label{th:opt_geom_conv}
	An update algorithm of Metropolis-Hastings type based on a \symker\ can not converge faster than geometrically.
\end{lemma}

\begin{proof}
	Let us revisit equation~\eqref{eq:p_of_v} in order to quantify the maximal reduction of the Lyapunov function, this time based on an arbitrary potential $V:\mathds{R}\rightarrow \mathds{R}$ (with normalisable induced probability density),
	\begin{align}
		(\mathcal{P}_AV)(z) &= \int_{-\infty}^{\infty}\md \gamma\mathcal{P}(z,z+\gamma)\left[A(z,z+\gamma)V(z+\gamma) + \left(1-A(z,z+\gamma)\right)V(z)\right]\\
		&= \int_{-\infty}^{\infty}\md \gamma\rho(\gamma)\left[V(z) + A(z,z+\gamma)\left(V(z+\gamma)-V(z)\right)\right]\\
		&= \int_{-\infty}^{0}\md \gamma\rho(\gamma)\left[V(z) + \left(V(z+\gamma)-V(z)\right)\right]\\
		&\quad + \int_{0}^{\infty}\md \gamma\rho(\gamma)\left[V(z) + \eto{-\left(V(z+\gamma)-V(z)\right)}\left(V(z+\gamma)-V(z)\right)\right]\\
		&= \int_{0}^{\infty}\md \gamma\rho(\gamma)\left[V(z) + V(z-\gamma) + \eto{-\left(V(z+\gamma)-V(z)\right)}\left(V(z+\gamma)-V(z)\right)\right]\label{eq:PV_integral}\\
		&\ge \int_{0}^{\infty}\md \gamma\rho(\gamma)V(z)\\
		&= \frac12 V(z)\,,
	\end{align}
	where we have used $V(z-\gamma)\ge0$ and as before $0\le\eto{-f(\gamma)}f(\gamma)$. Thus, the Lyapunov function cannot be reduced by more than a factor $\frac12$ by any single update step. In consequence, no convergence can be faster than the geometric reduction
	\begin{align}
		(\mathcal{P}_A^nV)(z) &\ge 2^{-n} V(z)
	\end{align}
	after $n$ steps.
\end{proof}

\begin{lemma}[Geometric drift, generalisation of \cref{th:exp_pot}]\label{th:geom_conv}
	A \nontriv\ \symker\ $\mathcal{P}$ together with a \mhacc\ satisfies the \sgdc\ for the potential $V:\mathds{R}^+\rightarrow \mathds{R}$ on $\mathds{R}^+\setminus [0,R]$ if $V$ fulfils
	\begin{align}
		\log V(z)-\log V(z') &\ge a(z-z')\label{eq:secant_exp_scaling}
	\end{align}
	for all $z\ge z'>R\ge0$ with some $a>0$.
\end{lemma}

\begin{proof}
	This is a generalisation of \cref{th:exp_pot}, which holds for exactly exponential potentials, for potentials that grow at least exponentially. In principle, this would suffice to end the proof here, but it is instructive to consider the full derivation starting from equation~\eqref{eq:PV_integral}
	\begin{align}
		(\mathcal{P}_AV)(z) &= \int_{0}^{\infty}\md \gamma\rho(\gamma)\left[V(z) + V(z-\gamma) + \eto{-\left(V(z+\gamma)-V(z)\right)}\left(V(z+\gamma)-V(z)\right)\right]\\
		&= \left[\frac12 + \int_{0}^{\infty}\md \gamma\rho(\gamma)\frac{V(z-\gamma)}{V(z)}\right] V(z) + K\\
		&= \left[\frac12 + \int_{0}^{\infty}\md \gamma\rho(\gamma)\eto{-\left(\log V(z) - \log V(z-\gamma)\right)}\right] V(z) + K\\
		&\le \left[\frac12 + \int_{0}^{\infty}\md \gamma\rho(\gamma)\eto{-a\gamma}\right] V(z) + K\\
		&= \alpha V(z) + K
	\end{align}
	with $\alpha<1$ and $0\le K\le \frac{1}{2e}$ as in equation~\eqref{eq:def_K}.
\end{proof}

\begin{lemma}[Arithmetic drift]\label{th:arith_conv}
	A \nontriv\ \symker\ $\mathcal{P}$ together with a \mhacc\ satisfies the \sadc\ for the potential $V:\mathds{R}^+\rightarrow \mathds{R}$ on $\mathds{R}^+\setminus [0,R]$ if $V$ fulfils
	\begin{align}
		V(z)-V(z') &\ge c(z-z')\label{eq:secant_lin_scaling}
	\end{align}
	for all $z\ge z'>R'\ge0$ with some $c>0$ and $R\ge R'$.
\end{lemma}

\begin{proof}
	Conceptually, this proof is very similar to the one of \cref{th:geom_conv}, but it has more subtleties. Again, we start with equation~\eqref{eq:PV_integral}
	\begin{align}
		(\mathcal{P}_AV)(z) &= \int_{0}^{\infty}\md \gamma\rho(\gamma)\left[V(z) + V(z-\gamma) + \eto{-\left(V(z+\gamma)-V(z)\right)}\left(V(z+\gamma)-V(z)\right)\right]\\
		&= V(z) + \int_{0}^{\infty}\md \gamma\rho(\gamma)\left[V(z-\gamma) - V(z)\right] + K\,,
	\end{align}
	but now we have to distinguish two cases.\footnote{There might be a more elegant proof that we could not find yet.} First, presume that for every $c>0$ there exists an $R'>0$ so that $V(z)-V(z') \ge c(z-z')$ for all $z\ge z'>R'\ge0$, i.e.\ asymptotically the potential grows strictly faster than linearly $V(z)\in \omega(z)$ (see tab.~\ref{tab:landau_symbols}). Then $V(z)-V(z-\gamma)$ becomes arbitrarily large for every $\gamma>0$. Since $\rho$ is \nontriv, it has non-vanishing overlap with the region $\gamma>\delta$ for some suitable $\delta>0$. Thus, the integral
	\begin{align}
		\int_{0}^{\infty}\md \gamma\rho(\gamma)\left[V(z) - V(z-\gamma)\right] &\ge \int_{\delta}^{\infty}\md \gamma\rho(\gamma)c\gamma
	\end{align}
	becomes arbitrarily large and in particular larger than $2K\le \frac{1}{e}$. With this we conclude
	\begin{align}
		(\mathcal{P}_AV)(z) &= V(z) - \int_{0}^{\infty}\md \gamma\rho(\gamma)\left[V(z) - V(z-\gamma)\right] + K\label{eq:PV_decomposed_arith}\\
		&\le V(z) -2K + K\\
		&= V(z) - K\,.
	\end{align}
	It remains to consider the case when there exists a $c>0$ so that for all $R'>0$ the potential growth is bound $V(z)-V(z') \le c(z-z')$, i.e.\ asymptotically the potential grows exactly linearly $V(z)\in\Theta(z)$ up to irrelevant local fluctuations. Then, asymptotically we can simplify $V(z)-V(z-\gamma)=V(z+\gamma)-V(z)=c\gamma$, which yields
	\begin{align}
		(\mathcal{P}_AV)(z) &= V(z) + \int_{0}^{\infty}\md \gamma\rho(\gamma)\left[-c\gamma + \eto{-c\gamma}c\gamma\right]\\
		&= V(z) - c\int_{0}^{\infty}\md \gamma\rho(\gamma)\gamma\left[1-\eto{-c\gamma}\right]\\
		&= V(z) - K'\,,
	\end{align}
	with a different constant
	\begin{align}
		K' &\coloneqq c\int_{0}^{\infty}\md \gamma\rho(\gamma)\gamma\left[1-\eto{-c\gamma}\right] >0\,.
	\end{align}
	Combining the cases of super-linear and strictly linear asymptotic growth of the potential $V$, completes the proof.
\end{proof}

\begin{lemma}[Diffusion]\label{th:diffus_conv}
	Every \nontriv\ \symker\ $\mathcal{P}$ together with a \mhacc\ satisfies the \ladc\ for every potential $V:\mathds{R}^+\rightarrow \mathds{R}$ (assuming $\eto{-V(z)}$ is normalisable) on $\mathds{R}^+\setminus [0,R]$ with some $R\ge0$.
\end{lemma}

\begin{proof}
	If $V$ satisfies the requirements of \cref{th:arith_conv}, then $\mathcal{P}$ already satisfies the \sadc\ which implies the \ladc\ because of \cref{th:drift_hierarchy}. It therefore remains to show the satisfaction of the \ladc\ in the case of strictly sub-linearly growing potentials. (Note that the potential still has to grow asymptotically in order to satisfy normalisability.) Sub-linear growth implies $V(z)-V(z-\gamma) \ge V(z+\gamma)-V(z)$ for all $\gamma>0$ and large enough $z>R$, therefore equation~\eqref{eq:PV_decomposed_arith} can be simplified
	\begin{align}
		(\mathcal{P}_AV)(z) &= V(z) - \int_{0}^{\infty}\md \gamma\rho(\gamma)\left[V(z) - V(z-\gamma)\right] + K\\
		&\le V(z) -\int_{0}^{\infty}\md \gamma\rho(\gamma)\left[V(z+\gamma) - V(z)\right] + K\\
		&= V(z) - \int_{0}^{\infty}\md \gamma\rho(\gamma)\left[\left(1- \eto{-\left(V(z+\gamma)-V(z)\right)}\right)\left(V(z+\gamma)-V(z)\right)\right]\\
		&= V(z) - K_z\,,
	\end{align}
	with
	\begin{align}
		K_z &\coloneqq \int_{0}^{\infty}\md \gamma\rho(\gamma)\left[\left(1- \eto{-\left(V(z+\gamma)-V(z)\right)}\right)\left(V(z+\gamma)-V(z)\right)\right]>0\,.
	\end{align}
	For any fixed $z$ the inequality $K_z>0$ holds and therefore, given a finite $z_0$, the minimum $K\equiv \min_{z\le z_0} K_z>0$ exists, as required by the \ladc. However, $K_z$ can become arbitrarily small which is why the \sadc\ is not satisfied in general.
\end{proof}

\begin{lemma}\label{th:min_growth}	
	$\begin{Bmatrix}
		\log V(z)\in\Omega\left(z\right) \\ V(z)\in\Omega\left(z\right)
	\end{Bmatrix}$
	as in tab.~\ref{tab:landau_symbols} implies
	$\begin{Bmatrix}
		\text{eq.~\eqref{eq:secant_exp_scaling}} \\ \text{eq.~\eqref{eq:secant_lin_scaling}}
	\end{Bmatrix}$
	in the setting of
	$\begin{Bmatrix}
		\text{\cref{th:geom_conv}} \\ \text{\cref{th:arith_conv}}
	\end{Bmatrix}$.
\end{lemma}

\begin{proof}
	Clearly, the linear scalings of $\log V(z)$ and $V(z)$ immediately satisfy the inequalities in \cref{th:geom_conv,th:arith_conv}, respectively. Since the potential grows at least with the given rate defined by the $\Omega(z)$, the differences $\log V(z)-\log V(z')$ and $V(z)-V(z')$ with $z>z'$ are at least as large as in the case of linear scaling.
\end{proof}

\begin{proof}[Proof: Classification of convergence types, \cref{th:classification}]
	This generalisation of \cref{th:dbl_exp_subst} follows from the respective \cref{th:geom_conv,th:arith_conv,th:diffus_conv} in the same way as \cref{th:dbl_exp_subst} follows as a multi-dimensional formulation of \cref{th:exp_pot}. Here, \cref{th:rates_of_conv} forms the link between drift condition and convergence while \cref{th:min_growth} guarantees that \cref{th:geom_conv,th:arith_conv,th:diffus_conv} are applicable, given the requirements to the potential. Moreover, \cref{th:compositions} implies that the combined algorithm satisfies the stronger version of the drift condition (in the hierarchy of \cref{th:drift_hierarchy}) because the \cdc-satisfying part has to satisfy at least the \wadc\ as well in order to allow for the required stationary distribution. 
	
	The no-go statement for super-exponential convergence follows directly from \cref{th:opt_geom_conv}.
\end{proof}

\begin{proof}[Proof: General HMC convergence, \cref{th:hmc_conv}]
	The \cdc\ is guaranteed by \cref{th:hmc_conv_compact} because $V$ is smooth. So it suffices to consider the drift condition.
	
	First, assume a very short trajectory length $T$ (the Langevin regime). Then the update proposed by the HMC is of size $\ordnung{T}$ (see tab.~\ref{tab:landau_symbols}) while the influence of the current position on this update is of order $\ordnung{T^2}$ and thus suppressed. In this regime the HMC becomes equivalent to an additive update by a multivariate normally distributed random number, that is the update becomes arbitrarily close to that of a \nontriv\ \symker. If the update step is small enough, the surface of the hypersphere $\Omega$ becomes locally flat and thus the update in radial direction (normal to $\Omega$) also approaches a \nontriv\ \symker. In this regime \cref{th:classification} is directly applicable.
	
	Now the molecular dynamics in longer trajectories introduce a drift of the configuration towards smaller potentials, so the drift that guarantees convergence can only increase and never decrease. This is also true for arbitrarily long trajectories (as used in~\cite{original_radial_update} to prove \cref{th:hmc_convergence_non-compact}) because the symplectic nature of the integrator imposes a fixed bound on the total energy and thus the potential $V$ independently of the trajectory length.
\end{proof}
 \end{document}